\documentclass[times,preprint,3p,sort&compress]{elsarticle}   
\usepackage{amssymb}
\usepackage{amsmath}

\usepackage{bm}         
\usepackage{amssymb}    
\usepackage{amsbsy}
\usepackage{amsmath}
\usepackage{amsfonts}
\usepackage{color}
\usepackage[dvipsnames]{xcolor}
\usepackage{amsthm}
\usepackage{caption}
\usepackage{subcaption}
\usepackage{amssymb}
\usepackage{enumitem} 
\setlist[itemize]{leftmargin=*}

\usepackage{xcolor}
\usepackage{graphicx}

\usepackage[hidelinks]{hyperref}
\hypersetup{
  colorlinks   = true, 	 
  urlcolor     = blue, 	 
  linkcolor    = blue, 	 
  citecolor   = red 	 
}

\DeclareMathOperator{\sech}{sech}

\DeclareMathOperator{\arccosh}{arccosh}

\begin{document}
\begin{frontmatter}

\title{Scattering between orthogonally wobbling kinks }
\author[salamanca1,salamanca2]{A. Alonso-Izquierdo}\ead{alonsoiz@usal.es}
\author[valladolid]{D.~Migu\'elez-Caballero}
\ead{david.miguelez@uva.es}
\author[valladolid]{L.M. Nieto}
\ead{luismiguel.nieto.calzada@uva.es}

\address[salamanca1]{Departamento de Mat\'ematica Aplicada, Universidad de Salamanca, Casas del Parque 2, 37008, Salamanca, Spain}
\address[salamanca2]{IUFFyM, Universidad de Salamanca, Plaza de la Merced 1, 37008, Salamanca, Spain}
\address[valladolid]{Departamento de F\'{\i}sica Te\'{o}rica, At\'{o}mica y \'{O}ptica
{ and Laboratory for Disruptive Interdisciplinary
Science (LaDIS)}, \\ 
Universidad de Valladolid, 47011 Valladolid, Spain}

\begin{abstract}
The resonant energy transfer mechanism, responsible for the presence of fractal patterns in the velocity diagrams of kink-antikink scattering, is analyzed for a family of two-component scalar field theory models, in which the kink solutions have two shape modes (one longitudinal and one orthogonal to the kink orbit), in addition to the zero mode, and in which energy redistribution can occur among these three discrete modes. 
We investigate the scattering between wobbling kinks whose orthogonal shape mode is initially excited, examining how the final velocities, amplitudes, and frequencies depend on the initial excitation amplitude. 
The differences that this model presents with respect to the $\phi^4$ model and its novel properties are highlighted.
This analysis sheds light on the intricate dynamics that arise from the interplay between multiple degrees of freedom in kink scattering processes, offering insights distinct from those observed in simpler models.
\end{abstract}

\begin{keyword}
Wobbling kinks, kink-antikink scattering, longitudinal and orthogonal modes, two-component scalar field theories. 
\end{keyword}

\end{frontmatter}

\section{Introduction}

The investigation of kink scattering has garnered significant attention in recent years due to its remarkable characteristics such as the presence of fractal structures in the velocity diagrams showing the final velocity of the scattered kinks as a function of the colliding velocity. Initially investigated in seminal references \cite{Campbell1983,Sugiyama1979,Anninos1991,Peyrard1983,Campbell1986}, the collision dynamics between kinks and antikinks in the $\phi^4$  model and deformed sine-Gordon models  unveiled two scattering channels: bion formation and kink reflection. Bion formation occurs when kinks collide and bounce back repeatedly, emitting radiation with each impact, while kink reflection involves a finite number of collisions before the kinks move apart. These channels dominate for different ranges of initial collision velocity, with bion formation prevalent at low velocities and kink reflection at higher velocities. A particularly intriguing phenomenon arises during the transition between these regimes, where bion formation and kink reflection intertwine infinitely, giving rise to a fractal structure within the velocity diagram. Understanding these intricate dynamics has significant implications for various physical applications, shedding light on nonlinear phenomena fostered by the presence of such topological defects \cite{Manton2004, Shnir2018, Rajaraman1982, Vilenkin2000, Vachaspati2006, Rebbi1984, Kevrekidis2019, Dauxois2006}.

The emergence of a fractal structure in the velocity diagram depicting kink scattering within the $\phi^4$ model stems from the presence of an internal vibrational mode, known as the shape mode, associated with kink solutions. 
This mode, together with the zero mode (or translational mode), allows the existence of the resonant energy transfer mechanism, which facilitates the redistribution of energy between kinetic and vibrational modes during kink collisions. 
Normally, in a scattering event, kinks and antikinks approach each other and collide, transferring some of the kinetic energy to the shape mode. 
This turns them into wobblers (vibrationally excited kinks) when they try to separate. 
If the kinetic energy of the wobblers is not enough, they can get closer and collide again. 
This cycle can continue indefinitely or cease after a finite number of collisions. In the latter scenario, enough vibrational energy is converted back to kinetic energy in the zero mode, enabling the wobblers to separate. 
This intricate mechanism and its associated phenomena have been widely investigated through various models, revealing profound complexity \cite{NavarroObregon2023,Adam2022,Adam2023,Long2024,Campos2023,Adam2019,Adam2021}. 
It is evident that the natural scenario in a physical system where topological defects can be found is that they are vibrating (i.e., with their shape modes excited), either due to collisions between them or due to the formation of topological defects after a phase transition. 
Therefore, from a physical point of view, the most interesting scenario lies in the investigation of scattering processes between wobbling kinks. 
The analysis of scattering between wobblers within the $\phi^4$ model has been widely discussed in previous works \cite{AlonsoIzquierdo2021b, AlonsoIzquierdo2022}. 
Through the numerical analysis of the scattering solutions derived from the Klein-Gordon partial differential equation, these investigations aim to shed light on the resonant energy transfer mechanism. It is worth noting that studies on wobblers scattering in the double sine-Gordon model have also been carried out in \cite{Gani2018,Gani2019}, contributing to a broader understanding of similar phenomena across different models \cite{AlonsoIzquierdo2019, Mohammadi2022,Halavanau2012, Dorey2011, Dorey2023, Bazeia2023, Campos2022, AlonsoIzquierdo2021, AlonsoIzquierdo2018, Belendryasova2019, AlonsoIzquierdo2020,Hahne2024,AlonsoIzquierdo2018b}. The resonant energy transfer mechanism has been recently demonstrated to occur in vortex scattering within the Abelian-Higgs model in (2+1)-dimensions, leading to the emergence of fractal patterns \cite{Krusch2024}. A collective coordinate model explaining this behavior is detailed in \cite{AlonsoIzquierdo2024b}. This discovery underscores the significance of investigating topological defects in (1+1)-dimensions, as many phenomena observed in this lower-dimensional context persist in higher dimensions. Notably, the phenomenon of spectral walls observed in kink dynamics \cite{Adam2019} has also been identified in vortex scattering \cite{AlonsoIzquierdo2024c}.

In the present work, our aim is to investigate the scattering between wobblers within a one-parametric family of relativistic (1+1)-dimensional two-component scalar field theories known as the MSTB model, which is a natural generalization of the $\phi^4$ model in this context. This system arises as a deformation of the $O(2)$ linear sigma model where the term breaking the symmetry includes the coupling constant $\sigma$. This model has been the focus of study by many researchers for decades \cite{Sarker1976, Subbaswamy1980b, Subbaswamy1980, Magyari1984, Ito1985, Ito1985b, Montonen1976, Rajaraman1975, AlonsoIzquierdo2008, Rajaraman1979, AlonsoIzquierdo2000}. 
The search for two-component kink solutions has been explored in references \cite{AlonsoIzquierdo2019,Sarker1976, Subbaswamy1980b, Subbaswamy1980, Magyari1984, Ito1985, Ito1985b}, while the generalization of such models is considered in \cite{Rajaraman1979, AlonsoIzquierdo2000,AlonsoIzquierdo2008}. 
These works have proved the existence of topological kinks and two families of non-topological kinks for $\sigma < 1$. 
However, when $\sigma \geq 1$, only topological kinks persist. The investigation of the quantum corrections to these kinks is discussed in \cite{Montonen1976, Rajaraman1975}, while the scattering behavior of these unexcited kinks is analyzed in \cite{AlonsoIzquierdo2019}.  
The existence of wobbling kinks in this model was described in \cite{AlonsoIzquierdo2023}, where it was shown that the simplest kink solutions that arise in this model have two shape modes, which allows their vibration in two different channels: one longitudinal and  another orthogonal to the kink orbit. 
Through perturbation theory, the radiation emission channels and the attenuation of the shape mode amplitudes were analytically identified. 
Therefore, the next natural step in studying the MSTB model is to investigate scattering between wobblers. In particular, we focus on the investigation of the scattering of wobbling kinks or wobblers emerging in the MSTB model in the regime $\sigma>1$. 
Several reasons justify the interest in carrying out this study. First, the MSTB model has been widely addressed in the literature, as mentioned earlier. Exploring the dynamics of topological defects in this model carries an implicit interest. 
Second, the study of scattering between wobblers has been limited to a few models, all of which involve a single scalar field. 
This restriction poses a limited scenario, exemplified by cases like the $\phi^4$ model, in which the kink possesses only one shape mode, causing the wobblers to vibrate longitudinally along their trajectory in internal space. 
However, 
the MSTB model stands out for its unique features: the kink exhibits two shape modes, allowing the wobblers to vibrate both longitudinally and orthogonally to their trajectory. 
This distinctive feature adds complexity and richness to the scattering dynamics in the MSTB model, warranting further exploration. This constitutes the fundamental reason for this research.
Note that now the resonant energy transfer mechanism can be activated, leading to energy redistribution among three discrete modes: the zero (or translational) mode (responsible for the kink motion), and the longitudinal and orthogonal shape modes (responsible for the kink vibration in each component of the field). 
This expansion in the number of modes involved introduces a new layer of complexity to the scattering dynamics, potentially yielding novel phenomena not observed in models with fewer degrees of freedom.

The organization of this paper is as follows. Section \ref{Sec:section2} introduces the MSTB model along with its kink solutions. The stability of these solutions and their vibrational modes are described. Section \ref{Sec:section3} outlines the numerical setup used for simulating collisions of wobbling kinks. Section \ref{Sec:section4} presents detailed numerical results, analyzing the behavior of the fractal patterns observed in the velocity diagrams resulting from the scattering between wobbling kinks, and exploring their dependency on both the coupling constant $\sigma$ and the initial excitation amplitude. By analyzing the final amplitudes and frequencies of the shape modes, the behavior of the resonant energy transfer mechanism in such models is examined. Finally, Section \ref{Sec:section5} provides the conclusions of the paper.

\section{The MSTB model: kink solutions, linear stability and shape modes}\label{Sec:section2}

In this paper we deal with a two-component scalar field theory model in a (1+1)-Minkowskian space-time, whose dynamics is governed by the Lagrangian density 
\begin{equation}\label{Eq:LagrangianDensity}
    \mathcal{L}=\frac{1}{2} \, \partial_\mu \phi \, \partial^{\mu} \phi+\frac{1}{2}\, \partial_\mu \psi \, \partial^\mu \psi- U(\phi, \psi),
\end{equation}
where the potential $U(\phi,\psi)$ is given by the fourth degree algebraic expression
\begin{equation}\label{Eq:Potential}
    U(\phi,\psi)=\frac{1}{2}(\phi^2+\psi^2-1)^2+\frac{\sigma^2}{2}\psi^2.
\end{equation}
In \eqref{Eq:LagrangianDensity} and \eqref{Eq:Potential}, $\phi$ and $\psi$ denote real scalar fields, and the Minkowski metric is represented as $g_{\mu,\nu}=\text{diag} \{ 1,-1\}$. Note that the potential term \eqref{Eq:Potential} introduces a coupling constant $\sigma^2$, which is a real positive parameter. This family of models is usually referred to as the MSTB model and has been extensively investigated in the literature, as indicated in the Introduction. Our focus here lies in analyzing the interaction between discrete modes (longitudinal and orthogonal) during collisions between topological defects in this model. The main objective of this paper is to understand how the mechanism of resonant energy transfer is influenced when the kinks include an orthogonal vibration channel. The reason for choosing the MSTB model for this study is twofold. The first is because it is a natural generalization of the $\phi^4$ model,  in which the energy transfer mechanism between modes was first studied. In fact, this two-component scalar field theory incorporates the $\phi^4$ model, which implies that the so-called $\phi^4$-kink arises as a solution of this model when the second field is null. This can be directly observed from the field equations of this model:
\begin{eqnarray} 
  \frac{\partial^2 \phi}{\partial t^2} -\frac{\partial^2 \phi}{\partial x^2} =2 \phi \left(1-\phi^2 -\psi^2\right), 
\qquad    \frac{\partial^2 \psi}{\partial t^2} -\frac{\partial^2 \psi}{\partial x^2} = 2 \psi \left(1-\phi^2 -\psi^2-\frac{\sigma^2}{2}\right). \label{Eq:FieldEqn2}
\end{eqnarray} 
It can be verified that the following static kink/antikink solutions satisfy the field equations (\ref{Eq:FieldEqn2}) of the model:
\begin{equation}\label{Eq:KinkSol}
    K^{(\pm)}(x)=(\phi(x),\psi(x))^\intercal=(\pm \tanh(x-x_0),0)^\intercal .
\end{equation}

\begin{figure}[htb]
    \centering
    \includegraphics[width=0.45\linewidth]{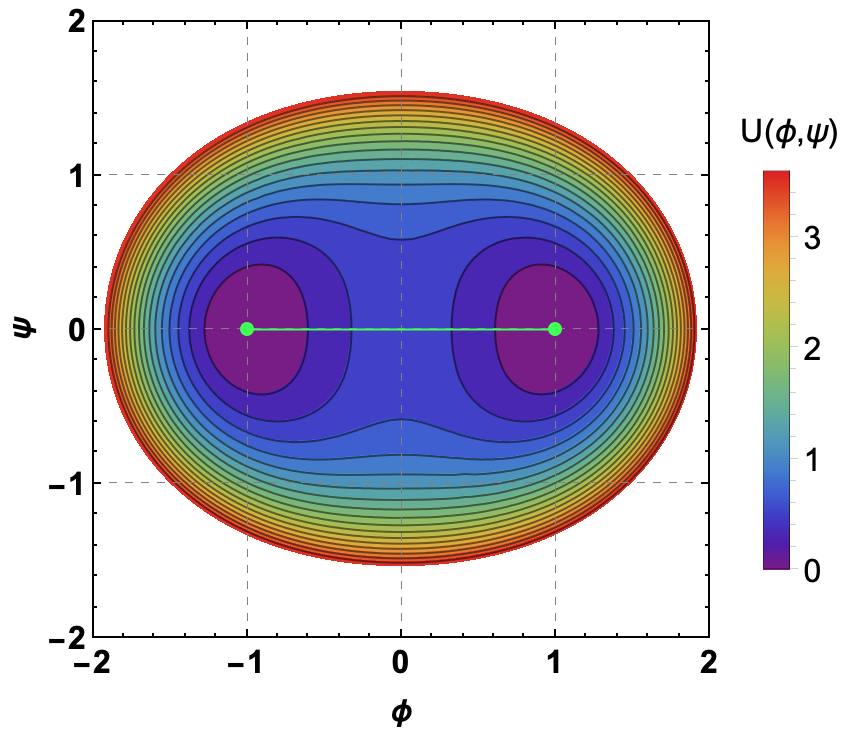}
    \caption{Contour density plot of the potential \eqref{Eq:Potential} for $\sigma=1.5$. The green line represents the orbit of the kink solutions \eqref{Eq:KinkSol}.}
    \label{Fig:PlotPotential}
\end{figure}
In Figure~\ref{Fig:PlotPotential}, the kink orbit has been depicted on the MSTB potential (\ref{Eq:Potential}). Note that both kink solutions (\ref{Eq:KinkSol}) interpolate between the two vacua or minimum points of \eqref{Eq:Potential}. For future use, it is  worth noting that these solutions can be transformed into traveling kinks 
\begin{equation}\label{Eq:KinkSolLorentz}
    K^{(\pm)}(x,t)=\left(\pm \tanh \frac{x-x_0-v_0 t}{\sqrt{1-v_0^2}} \, , \, 0\right)^\intercal,
\end{equation}
simply by applying a Lorentz boost, where $v_0$ is the kink velocity and $-1<v_0<1$. 
The second reason that makes the MSTB model a natural candidate to develop the study we have embarked on is that the kink solution (\ref{Eq:KinkSol}) can vibrate both longitudinally and orthogonally with respect the kink orbit, and these channels include only one eigenstate each. 
Specifically, the second-order small kink fluctuation operator involves only one  longitudinal vibration mode (the so-called shape mode, already present in the $\phi^4$ model).
But we also have an orthogonal vibration mode, that is, as it will be explained later, the second order fluctuation operator consists of two decoupled Schr\"odinger-type equations whose corresponding resolution will provide us with a single mode shape each.
This scenario provides a natural framework to analyze the energy transfer mechanism between longitudinal and orthogonal modes. This statement is justified in the following way:  substituting a linearly perturbed kink
\begin{equation}\label{Eq:KinkPlusPertubation}
    \widetilde{K}(x,t;\omega, a)=K^{(\pm)}(x)+ a\, e^{i \omega t} F(x)=
    \begin{pmatrix}
\pm \tanh(x)  \\ 
0
    \end{pmatrix}
    + a\, e^{i \omega t}
    \begin{pmatrix}
\overline{\eta}(x) \\ 
\widehat{\eta}(x)
    \end{pmatrix}
    ,
\end{equation}
into the field equations  \eqref{Eq:FieldEqn2} leads to the spectral problem
\begin{equation}\label{Eq:SpectralProblem}
    \mathcal{H}
\begin{pmatrix}
\overline{\eta} (x)  \\ 
\widehat{\eta} (x)
    \end{pmatrix}    = 
    \omega^2 
\begin{pmatrix}
\overline{\eta} (x)  \\ 
\widehat{\eta} (x)
    \end{pmatrix}    ,
\end{equation}
where ${\cal H}$ is given by the matrix differential operator
\begin{equation}\label{Eq:HMatrix}
    \mathcal{H}=
    \begin{pmatrix}
    -\dfrac{d^2}{d x^2}+ 4-6 \sech^2 x  & 0  \\
    0 &-\dfrac{d^2}{d x^2}  +\sigma^2 -2 \sech^2 x  \\
    \end{pmatrix}
    .
\end{equation}
From \eqref{Eq:HMatrix} it is clear that the longitudinal modes (vibrating along the kink orbit) 
\begin{equation}\label{Eq:LongFluc}
\overline{F}_{\omega}(x)=
    \begin{pmatrix}
\overline{\eta} (x)  \\ 
0
    \end{pmatrix} 
    ,
\end{equation}
are decoupled from the orthogonal modes (vibrating orthogonally with respect the kink orbit)
\begin{equation}\label{Eq:OrthoFluc}
\widehat{F}_{\omega}(x)=
    \begin{pmatrix}
 0 \\ 
\widehat{\eta} (x) 
    \end{pmatrix} 
    .
\end{equation}
In addition to this, the longitudinal and orthogonal components $\overline{\eta} (x)$ and $\widehat{\eta} (x)$ are respectively determined as the eigenfunctions of the Schr\"odinger-like operators 
\[
\mathcal{H}_{11}=-\dfrac{d^2}{d x^2}+ 4-6 \sech^2 x \hspace{0.5cm} \mbox{and} \hspace{0.5cm} \mathcal{H}_{22}=-\dfrac{d^2}{d x^2}  +\sigma^2 -2 \sech^2 x\, ,
\]
 with P\"oschl-Teller potential wells  \cite{Flugge1971,Morse1953,Morse1933}. Therefore, the eigenfunctions associated with the operator \eqref{Eq:HMatrix} are described as:
\begin{itemize}
\item \textit{Longitudinal eigenmodes:} The fluctuations of the kink solution correspond to those found in the $\phi^4$ model with a single component \cite{Vachaspati2006, Shnir2018, Manton1997, Barashenkov2009, Barashenkov2009b, AlonsoIzquierdo2023}. Therefore, the discrete eigenfrequencies are given by $\omega=0$ and $\overline{\omega}=\sqrt{3}$, whose eigenfunctions are respectively the \textit{zero mode}  
\begin{equation}\label{Eq:ZeroMode}
     \overline{F}_{0} (x) =        
     \begin{pmatrix}
    \overline{\eta}_0 (x) \\
    0
    \end{pmatrix}
= \begin{pmatrix}
    \sech^2 x \\
    0
    \end{pmatrix},
\end{equation}
and the \textit{longitudinal shape mode}
\begin{equation}\label{Eq:LongitudinalShapeMode}
     \overline{F}_{\sqrt{3}} (x) =        
     \begin{pmatrix}
    \overline{\eta}(x) \\
    0
    \end{pmatrix}
= \begin{pmatrix}
    \sech x \tanh x \\
    0
    \end{pmatrix}.
\end{equation}
Additionally, a continuous spectrum arises on the threshold value $\overline{\omega}_{0}^c=2$ with frequency $\overline{\omega}_{\overline{q}}^c=\sqrt{4+\overline{q}^2}$, where $\overline{q}$ is a real positive parameter. The corresponding eigenfunctions for this case can be written as
    \begin{equation}\label{Eq:ContinuousLongitudinalMode}
        \overline{F}_{\sqrt{4+\bar{q}^2}} \,(x)=
        \begin{pmatrix}
    \overline{\eta}_{\bar{q}}(x) \\
    0
    \end{pmatrix} 
=
    \begin{pmatrix}
    ( -1-\bar{q}^2+ 3 \, \tanh^2 x -3 i \bar{q} \tanh x) \, e^{i \bar{q} x} \\
    0
    \end{pmatrix} . 
    \end{equation}
\item \textit{Orthogonal eigenmodes:} Only one discrete mode appears, whose frequency $\widehat{\omega}=\sqrt{\sigma^2-1}$ depends on the coupling constant $\sigma$, while the now-called \textit{orthogonal shape mode} is determined as
\begin{equation}\label{Eq:OrthogonalsShapeMode}
     \widehat{F}_{\sqrt{\sigma^2-1}} (x) =        
     \begin{pmatrix}
    0 \\
    \widehat{\eta} (x)
    \end{pmatrix}
= \begin{pmatrix}
    0 \\
     \sech x
    \end{pmatrix}.
\end{equation}
Note that these fluctuations describe perturbations which are orthogonal to the kink orbit, in the direction of the second field $\psi$. It is clear from the expression of the eigenvalue $\widehat{\omega}$ that the kink (\ref{Eq:KinkSol}) is unstable when $\sigma < 1$. 
Consequently, a less energetic kink arises in the theory in the same topological sector. In this regime the kink (\ref{Eq:KinkSol}) decays into this new kink, which takes the form \cite{AlonsoIzquierdo2019, AlonsoIzquierdo2023}
\begin{equation*}
    K^*(x)=(\pm q \tanh(\sigma(x-x_0)),\lambda \sqrt{1-\sigma^2}\sech(\sigma(x-x_0)))^\intercal,
\end{equation*}
where $q,\lambda=\pm1$.
However, for the regime $\sigma > 1$, the kink is stable and the orthogonal shape mode corresponds to a fluctuation in the second field (which  was initially zero), describing a vibration of the kink orbit, which we can visualize like the movement of a violin string.
For this reason, the scattering of orthogonally wobbling kinks addressed in this article will  be restricted to the range $\sigma>1$. 
Along with the discrete mode (\ref{Eq:OrthogonalsShapeMode}) there is also a continuous spectrum with eigenfunctions 
\begin{equation}\label{Eq:ContnuousOrthogonalMode}
     \widehat{F}_{\sqrt{\sigma^2-1}} (x) =        
     \begin{pmatrix}
    0 \\
    \widehat{\eta}_{\widehat{q}} (x)
    \end{pmatrix}
= \begin{pmatrix}
    0 \\
    (\,\widehat{q}+ i \tanh x)\, e^{i \widehat{q} x}
    \end{pmatrix}, \qquad \widehat{\omega}_{\widehat{q}}^c=\sqrt{\sigma^2+\widehat{q}^2}.
\end{equation}
\end{itemize}

\begin{figure}[htb]
         \centering
         \includegraphics[width=0.45\textwidth]{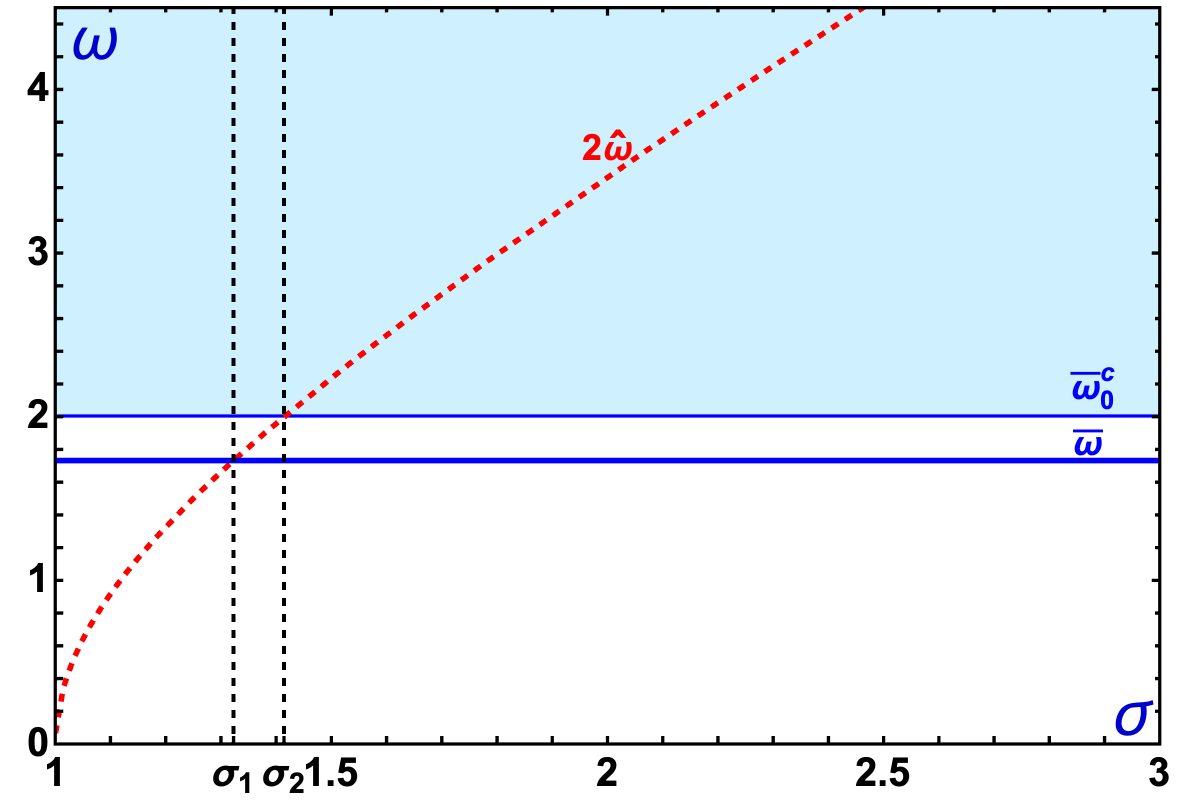}
         \hfill
         \includegraphics[width=0.45\textwidth]{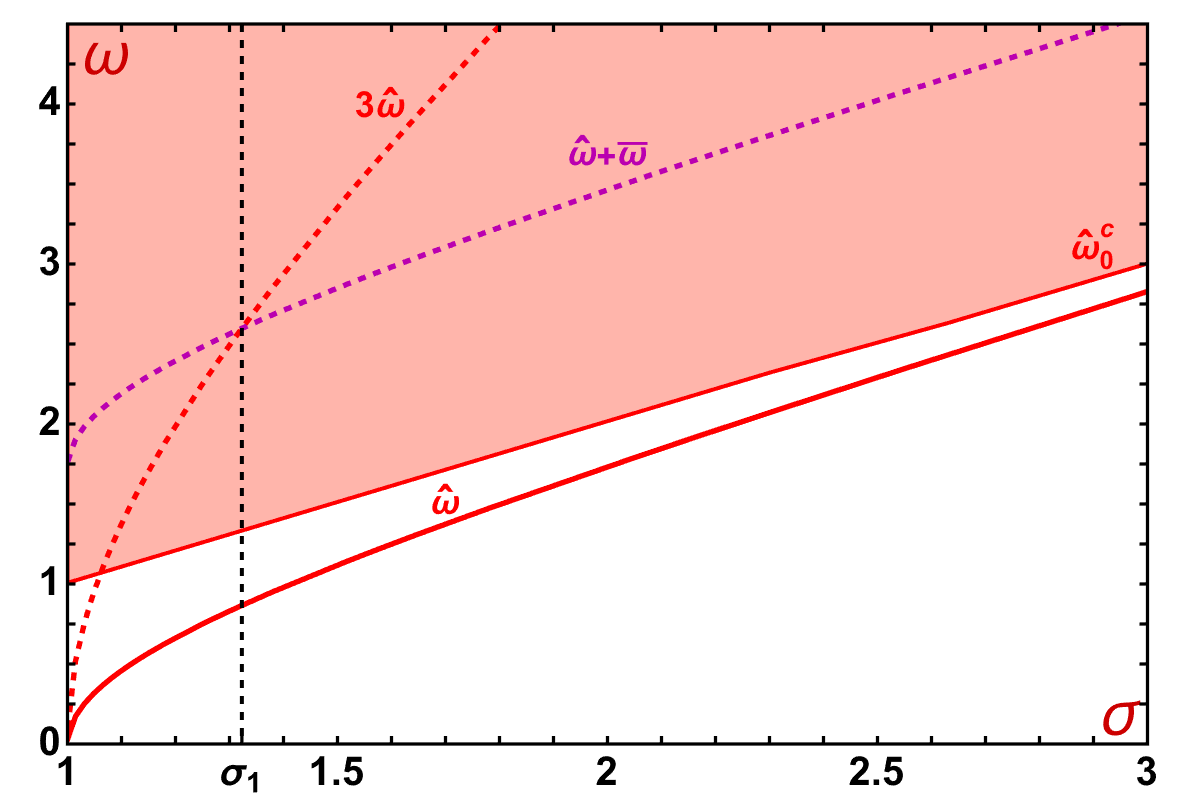}
\caption{Spectrum of the longitudinal (left) and orthogonal (right) fluctuations of the operator \eqref{Eq:HMatrix}.
The blue and red areas coincide with the continuous eigenvalues for the first and second field components. The dashed colored lines represent the radiation frequencies that arise when $\widehat{\eta}$ is initially triggered, while the black dashed vertical lines represent the values at which there is resonance between two frequencies.}
     \label{Fig:VibrationSpectrum}
\end{figure}
The spectrum of the kink fluctuation operator (\ref{Eq:HMatrix}) as a function of the coupling constant $\sigma$ is shown in Figure~\ref{Fig:VibrationSpectrum}, in which the graph on the left represents the spectrum of longitudinal fluctuations, while the one on the right corresponds to orthogonal fluctuations. The discrete eigenfrequencies are represented by solid lines, while the continuous spectrum appears as a shaded region. 
Note that the longitudinal spectrum (left plot) does not depend on $\sigma$, exhibiting a single longitudinal eigenfrequency denoted as $\overline{\omega}=\sqrt{3}$ and a continuous spectrum emerging in the threshold value $\overline{\omega}_0^c= 2$. 
On the other hand, the orthogonal spectrum changes depending on the value of $\sigma$, presenting a single orthogonal frequency $\widehat{\omega}=\sqrt{\sigma^2 -1 }$, and a continuum that emerges from the value $\widehat{\omega}_0^c = \sigma$. 
As a first conclusion, it should be noted that the orthogonal shape mode is easily excited when the coupling constant is close to the value 1. 
In particular, it can be observed that in the range $\sigma\in [1,2]$, the orthogonal eigenfrequency is lower than the longitudinal one and, therefore, more prone to excitation. 
When $\sigma >2$, the situation is reversed and it will be easier to excite the longitudinal mode. 
Since these regimes are expected to lead to different collision behavior of the wobblers, simulations with  values of $\sigma$  both lower and higher than $2$ (typically the values $\sigma=1.2$, $\sqrt{7}/2$, $\sqrt{2} $, $1.5$ and $ 2.5$) will be carried out in Section~\ref{Sec:section4}.
Additionally, certain harmonics of the discrete eigenfrequencies are represented by dashed curves in Figure~\ref{Fig:VibrationSpectrum}. These frequencies can be excited because of nonlinear terms of the theory at higher orders and play an essential role in radiation phenomena. This type of process where a kink is initially excited by the orthogonal shape mode $\widehat{\eta}$ has been analyzed in \cite{AlonsoIzquierdo2023} by means of a perturbation approach. It was found that the orthogonally wobbling kink is able to emit radiation with frequency $2\widehat{\omega}$ in the longitudinal channel.  Note that the longitudinal radiation captures the information coming from the orthogonal channel. On the other hand, orthogonal radiation (traveling in the second field) is emitted with frequencies $3\widehat{\omega}$ and $\widehat{\omega}+\overline{\omega}$. Typically, the radiation emission remains small when the initial amplitude of the orthogonal shape mode is reasonably small. However, for certain values of the coupling constant $\sigma$, resonance phenomena occur in which this radiation emission can increase by one or two orders of magnitude. These special model parameter values can be visualized in Figure~\ref{Fig:VibrationSpectrum} by the coincidence of certain eigenfrequencies, and are listed below:
\begin{itemize}
    \item 
    $\mathbf{\sigma_1=\frac{\sqrt{7}}{2}} \approx 1.32288$\textbf{:}
    As seen in Figure~\ref{Fig:VibrationSpectrum} (left), a resonance occurs when twice the eigenfrequency of the orthogonal shape mode $2\widehat{\omega}$ coincides with the longitudinal discrete eigenfrequency $\overline{\omega}$ for the model parameter $\sigma_1$. 
    In this case,  a second resonance coincidentally arises between $3\widehat{\omega}$ and $\widehat{\omega}+\overline{\omega}$ in the orthogonal fluctuation spectrum, see Figure~\ref{Fig:VibrationSpectrum} (right). 
    In this scenario, there is a significant loss of energy from the orthogonal shape mode in favor of the longitudinal mode, accompanied by a pronounced emission of radiation in the orthogonal channel with frequencies $\overline{\omega} + \widehat{\omega}$ and $3\widehat{\omega}$ \cite{AlonsoIzquierdo2023}.  
    Obviously, this phenomenon can have relevant effects on the resonant energy transfer mechanism, causing wobbler collisions to exhibit singular behaviors within the values of the coupling constant $\sigma$. In other words, as the wobblers approach each other, a significant portion of the energy from the orthogonal shape mode is transferred to the longitudinal shape mode  and also radiated away. Consequently, for this $\sigma$ value, the collision dynamics is predominantly governed by the scattering of longitudinally wobbling kinks. 
    Under these circumstances, we can anticipate that behavior is similar to the wobbler scattering observed in the $\phi^4$ model.
    Deviations from the patterns observed  in this case provide valuable information about the interaction between the orthogonal shape mode and other modes within the resonant energy transfer mechanism.
    
    \item 
    $\mathbf{\sigma_2=\sqrt{2}} \approx 1.41421$\textbf{:} 
    In this case, the frequency $2\widehat{\omega}$ begins to form part of the continuous longitudinal spectrum (see Figure~\ref{Fig:VibrationSpectrum} left). In fact, this value is characterized by the coincidence between the values $2\widehat{\omega}$ and $\overline{\omega}_0^c$. This implies that the quadratic longitudinal radiation source is activated from this point. 
    As a consequence, the radiation emission in the longitudinal channel with frequency $2\widehat{\omega}$ is very significant (relative to other values of $\sigma$). 
    This implies that a large amount of energy is extracted from the orthogonal shape mode in the form of longitudinal radiation. 
Again, in this situation some peculiar behaviors could appear in the scattering between wobblers.    However, despite this phenomenon, the energy loss resulting from radiation emission is not considered significant enough to alter the wobbler scattering characteristics for this particular value.
\end{itemize}

In the analysis of kink scattering in this model, we will take into account previous observations regarding the relevant values of the coupling constant $\sigma$, with the aim of performing simulations at various values of the model parameter that can lead to different behaviors. 
We have already mentioned the values $\sigma=1.5$ and $\sigma=2.5$ as representative of possible different behaviors. 
To these two values we will also add $\sigma=1.2$, for which the emission of radiation in the longitudinal channel with frequency $2\widehat{\omega}$ is suppressed. 
Additionally, studies of the scattering of orthogonally wobbling kinks will be carried out in the special cases $\sigma_1=\sqrt{7}/2$ and $\sigma_2=\sqrt{2}$ mentioned above.

\section{Numerical setup for the scattering between wobblers in the MSTB model}\label{Sec:section3}

In this paper we are analyzing  collisions between a kink and antikink \eqref{Eq:KinkSol}, with their orthogonal shape modes initially activated. 
It is well-known that the scattering between wobbling kinks in the $\phi^4$ model exhibits complex fractal structures depending on the vibration amplitude of its only shape mode \cite{AlonsoIzquierdo2021b,AlonsoIzquierdo2022}. 
These fractal structures arise due to the resonant energy transfer mechanism between the zero and shape discrete modes. In this model both of these eigenfluctuations are embedded in the longitudinal channel. 
Our goal now is to numerically investigate whether this is a necessary condition to trigger the resonant energy transfer mechanism or whether it can also be initiated by orthogonal eigenmodes. 
The initial numerical setup will consist of a well-separated orthogonally wobbling kink/antikink configuration whose centers are initially located at points $x_0$ and $-x_0$ (with $x_0 \gg 1$) and travel respectively with velocities $v_0$ and $-v_0$.
Therefore, the initial configuration for the numerical simulations can be characterized as 
\begin{equation}\label{Eq:InitialConditionSimulation}
   K^{(\pm)}_{W K}(x,t,v_0,x_0,\sigma,a_0) \cup K^{(\mp)}_{WK}(x,t,-v_0,-x_0,\sigma,a_0)=\left\lbrace
   \begin{array}{cc}
       K^{(\pm)}\left(  \Tilde{x}_{K^{\pm}}\right)+\widehat{a}_0 \, \sin (\widehat{\omega} t)\, \widehat{F}_{\widehat{\omega}} \left(  \Tilde{x}_{K^{\pm}}\right) , \quad  \mathrm{if} \, x<0, \vspace{0.3cm} \\ 
        K^{(\mp)}\left(\Tilde{x}_{K^{\mp}}\right)+ \widehat{a}_0\, \sin( \widehat{\omega} t)\, \widehat{F}_{\widehat{\omega}} \left(\Tilde{x}_{K^{\mp}}\right) ,  \quad \mathrm{if}\, x>0,
   \end{array}
   \right.
\end{equation}
where
\begin{equation*}
    \Tilde{x}_{K^{\pm}}=\frac{x+x_0-v_0 t}{\sqrt{1-v_0^2}}, \qquad \Tilde{x}_{K^{\mp}}=\frac{x-x_0+v_0 t}{\sqrt{1-v_0^2}}. 
\end{equation*}
Here, $\widehat{a}_0$ is the initial amplitude of the orthogonal shape mode, which measures the excitation of the wobbling kinks. 
Note that \eqref{Eq:InitialConditionSimulation} determines well-behaved initial conditions because the single kinks in this model are exponentially localized. 
It is expected that during the scattering process there will be an energy transfer between the orthogonal, longitudinal and translational modes, together with a small radiation emission in both field components. 
This energy loss causes  the wobbling amplitude $\widehat{a}$ to decay following the theoretical law
\begin{equation}\label{Eq:DecayLaw}
    \widehat{a}(t)^2\approx \frac{\widehat{a}(0)^2}{1+ \widehat{\omega}  \, \, \widehat{c}\,\,  \widehat{a}(0)^2\, t},
\end{equation}
where $\widehat{c}$ is a constant that  depends on the radiation amplitudes and $\widehat{\omega}$ and, therefore, also on $\sigma$ \cite{AlonsoIzquierdo2023}. Generally, this phenomenon only plays an important  role when $\widehat{a}(0)>0.1$. In a scattering event, once the collision occurs, a part of the energy could be transferred to the longitudinal eigenmode. 
In this case, the resulting longitudinally wobbling kinks behave approximately  like those of  $\phi^4$, that is, the kink would emit radiation in the longitudinal channel whose frequency would  be $2\overline{\omega}$. 
If this happens, the wobbling amplitude of this internal mode will also decay  following the formula 
\begin{equation}
        \overline{a}(t)^2\approx \frac{\overline{a}(0)^2}{1+ \overline{\omega} \, \,\overline{c} \,\,  \overline{a}(0)^2\, t},
\end{equation}
where $\overline{c}$ is a constant related to the radiation amplitude \cite{Manton1997, Barashenkov2009,Barashenkov2009b}. 
As before, this decrease is only relevant only for amplitudes $\overline{a}(0) > 0.1$. This phenomenon implies that the initial amplitudes of the shape modes of the colliding kinks will decrease as they approach each other. 
When the initial amplitudes of the  shape modes involved are small, the described phenomenon can be considered negligible and these magnitudes are suitable parameters to describe the scattering processes described in this article.

Based on all the results presented above, after the collision of the wobbling kinks and in the case that they are reflected, it is expected that the kinks will move away with a certain final velocity $v_f$, with both longitudinal and orthogonal shape modes excited, that is, with a configuration of the form:
\begin{small}
\begin{equation}
   K^{(\pm)}_{W K,f}(x,t,v_0,x_0,\sigma,a_0) \cup K^{(\mp)}_{WK,f}(x,t,-v_0,-x_0,\sigma,a_0)=\left\lbrace
 \!\!  \begin{array}{cc}
       K^{(\pm)}\left(  \Tilde{x}_{K^{\pm},f}\right)+\overline{a}_f  \sin (\overline{\omega} t+\delta)\, \overline{F}_{\overline{\omega}}\left(  \Tilde{x}_{K^{\pm},f}\right)+\widehat{a}_f  \sin (\widehat{\omega} t)\, \widehat{F}_{\widehat{\omega}} \left(  \Tilde{x}_{K^{\pm},f}\right) , \,   \mathrm{if} \, x<0, \vspace{0.3cm} \\ 
        K^{(\mp)}\left(\Tilde{x}_{K^{\mp},f}\right)+\overline{a}_f \sin (\overline{\omega} t+\delta)\, \overline{F}_{\overline{\omega}} \left(  \Tilde{x}_{K^{\mp},f}\right)+\widehat{a}_f  \sin (\widehat{\omega} t)\, \widehat{F}_{\widehat{\omega}} \left(  \Tilde{x}_{K^{\mp},f}\right)  ,  \,  \mathrm{if}\, x>0,
   \end{array}
   \right.
\end{equation}
\end{small}
where
\begin{equation}\label{Eq:InitialConditionSimulation2}
    \Tilde{x}_{K^{\pm},f}=\frac{x+v_f t}{\sqrt{1-v_f^2}}, \qquad \Tilde{x}_{K^{\mp},f}=\frac{x-v_f t}{\sqrt{1-v_f^2}}. 
\end{equation}

 Once the initial conditions of our problem have been described, the kink scattering processes are simulated by discretizing equations \eqref{Eq:FieldEqn2} using a fourth-order algorithm designed specifically  to handle general nonlinear systems of two coupled Klein-Gordon partial differential equations, as described in \cite{AlonsoIzquierdo2021}. 
 To prevent the radiation emitted by the system from being reflected from the spatial grid boundaries, Mur absorbing boundary conditions were implemented at $x=100$ and $x=-100$. 
 Simulations were performed for various values of the coupling constant $\sigma \in [1.2,2.5]$ over a range of initial velocities $v_0\in\left[0.1, 0.9\right]$ with increments of $\Delta v_0=10^{-5}$. For velocity ranges where the kinks experience only a rebound, the step size was increased to $\Delta v_0=10^{-3}$. 
 In each of these cases the simulations have been carried out changing the initial orthogonal amplitude $\widehat{a}_0$ with steps $\Delta \widehat{a}_0 = 0.02$ in the range $\widehat{a}_0 \in (0,0.2)$. 
 The total number of simulations carried out to obtain the results presented in this article amounts to approximately $10^6$, which have been carried out by using the \textit{Cal\'endula} Supercomputer integrated into the high-performance computing resources of the Castilla y Le\'on Supercomputing Center (SCAYLE). All simulations were performed within the spatial interval $x\in[-100,100]$ with a step size of $\Delta x=0.005$. 
 Initially, the kink/antikink centers are separated by a distance $d=2x_0=15$. 
 When kinks are reflected, their final  velocities are measured numerically. In addition to this, the amplitudes of the orthogonal and longitudinal shape modes are also estimated by using a fast Fourier transform algorithm, implemented at the comoving kink centers $\widehat{x}_M=x_C$ and the respective points
\begin{equation*}
    \overline{x}_M=x_C\pm \sqrt{1-v_0^2} \, \arccosh\sqrt{2},
\end{equation*}
where the discrete longitudinal fluctuations present their maximum elongation.

It is also  worth mentioning that the initial conditions proposed in \eqref{Eq:InitialConditionSimulation} are symmetric with respect to the origin of the spatial coordinate. 
That is, the kink and the antikink approach each other at the same velocity (this does not imply a loss of generality since it is simply assumed that we fix the reference system with the origin located at the center of mass) and their shape modes have been excited with the same amplitude and phase. 
As demonstrated in \cite{AlonsoIzquierdo2023, AlonsoIzquierdo2024}, this scenario provides the most extreme phenomena that can arise in kink collisions, since at the moment of impact a constructive interference occurs, amplifying the redistribution of energy between the  vibrational and translational modes. 
Since the numerical algorithm used to discretize the field equations (\ref{Eq:FieldEqn2}) also preserves this symmetry, the scattering results can be extracted directly from one of the traveling wobblers involved in the simulations.

\section{Numerical results for the scattering between wobblers in the MSTB model}\label{Sec:section4}

Now that we have detailed the implementation of the numerical simulations, we proceed to present the main results derived from the data obtained from the simulations. 
The parameter $\sigma$ emerges as pivotal to understanding the energy transfer between internal modes and significantly influences the velocity diagrams of the scattering process. In fact, $\sigma$ plays a crucial role in the appearance of new single-bounce windows in these diagrams. Indeed, as it will be explained later, increasing the value of $\sigma$ is equivalent to increasing the energy needed to trigger the orthogonal eigenmode. This implies that, for example, for the same value of  $\widehat{a}_0$ the critical velocity where the one-bounce tail arises will be smaller as $\sigma$ is increased,  since more energy  can be transferred to the translational eigenmode.
Furthermore, the amplitude of the orthogonal shape mode holds considerable importance: a larger $\widehat{a}_0$ facilitates greater energy transfer to the longitudinal shape mode and the translational one, facilitating separation of the wobblers. The main characteristics of these scattering processes are delineated in the following subsections. 
Firstly, Section \ref{Sec:section4.1} delves into the analysis of the velocity diagrams, elucidating their behavior in each of the cases. 
Next, in Section \ref{Sec:section4.2}, we investigate the resonant energy transfer mechanism by examining the final wobbling amplitudes (both longitudinal and orthogonal), which provides information on the strength of shape mode excitations at the end of the process.

\subsection{Properties of the velocity diagrams for the scattering between wobblers in the MSTB model}\label{Sec:section4.1}

In this subsection we will describe the characteristics of the velocity diagrams observed in the scattering between wobblers within the MSTB model, when the orthogonal shape mode is initially excited. We will classify these scenarios according to the strength of the initial excitation into three distinct regimes. 
We will first examine the regime in which wobblers are weakly excited, typically for $\widehat{a}_0<0.05$. 
Next, we will investigate the regime characterized by moderate excitations, within the range $0.05<\widehat{a}_0<0.1$. 
Finally, we will explore the strong excitation regime, identified by $\widehat{a}_0>0.1$. 
Through this analysis we aim to provide a comprehensive understanding of the behavior exhibited by the velocity diagrams (and the fractal structures included in them) when varying the value of the initial excitation and also the coupling constant $\sigma$, as we have already indicated in Section \ref{Sec:section2}.

\begin{figure}[!ht]
\centering
\begin{subfigure}{1\textwidth}
    \includegraphics[width=1.0\linewidth]{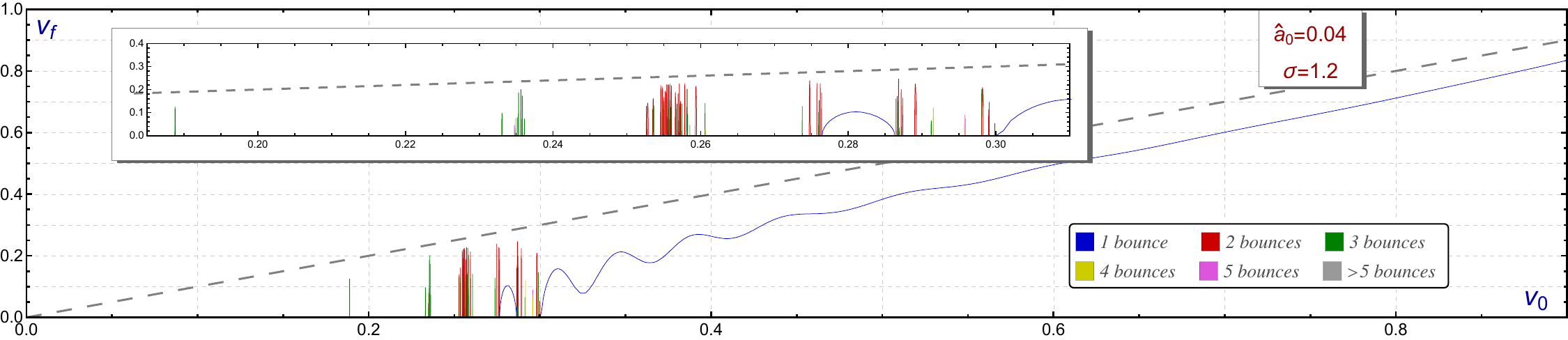}
\end{subfigure}
\hfill
\begin{subfigure}{1\textwidth}    \includegraphics[width=1.0\linewidth]{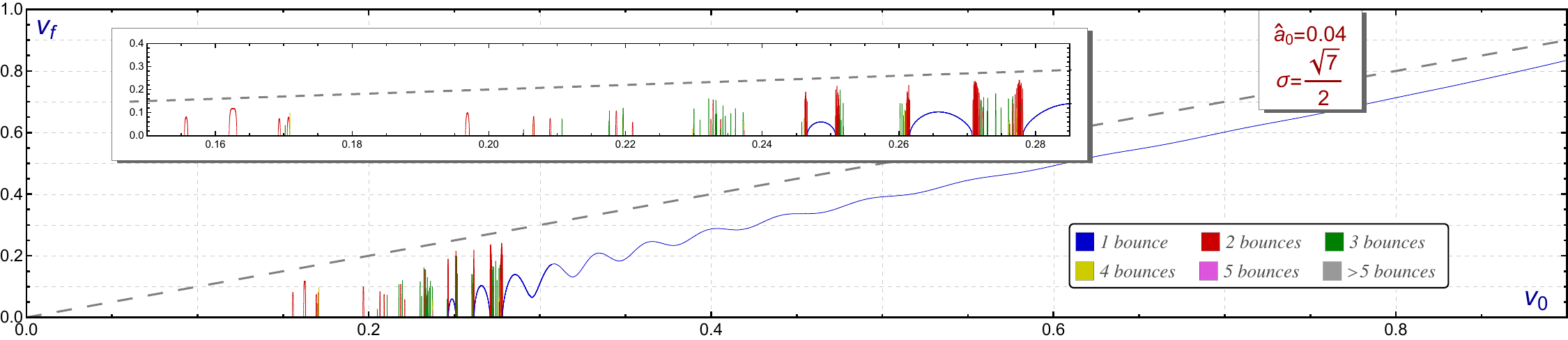}
\end{subfigure}
\hfill
\begin{subfigure}{1\textwidth}
    \includegraphics[width=1.0\linewidth]{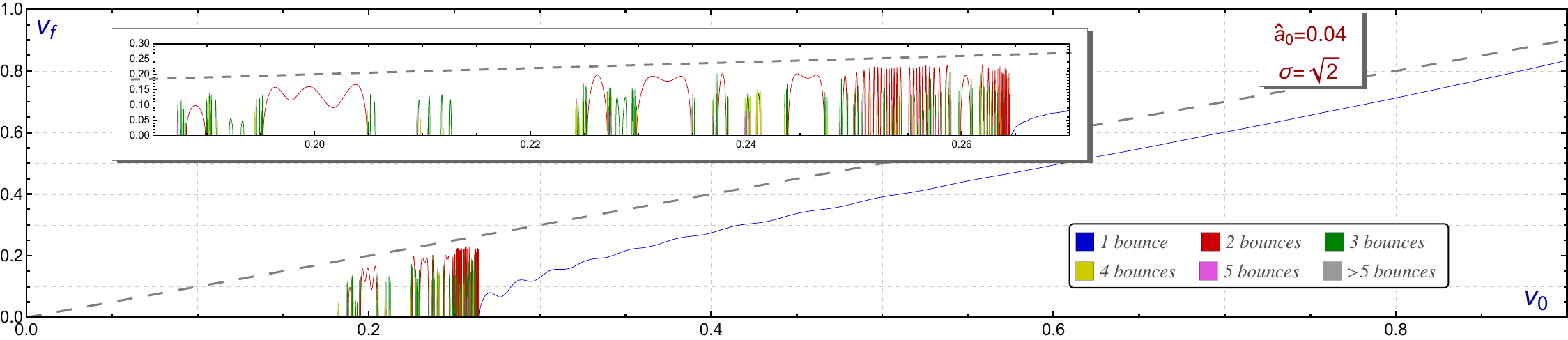}
\end{subfigure}
\hfill
\begin{subfigure}{1\textwidth}
    \includegraphics[width=1.0\linewidth]{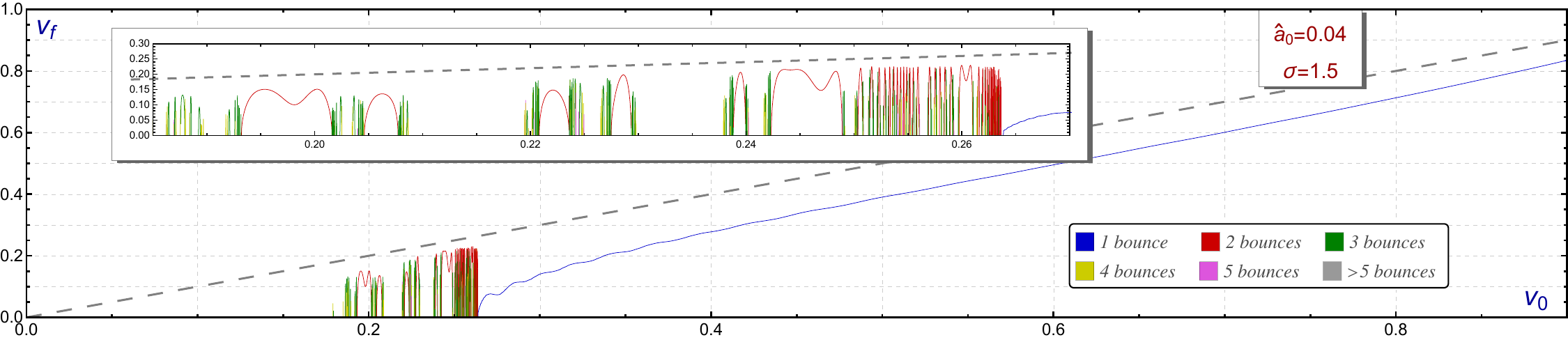}
\end{subfigure}
\hfill
\begin{subfigure}{1\textwidth}
    \includegraphics[width=1.0\linewidth]{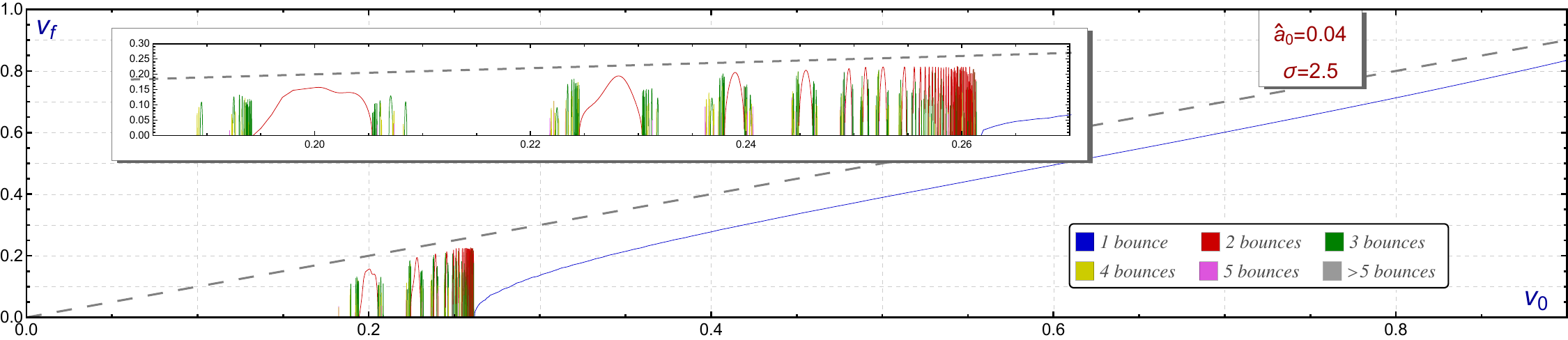}
\end{subfigure}
        
  \caption{Final velocity $v_f$ of the scattered kinks as a function of the initial velocity $v_0$. The color code shown in the graphs indicates the number of bounces suffered by the kink-antikink pair before moving appart. 
  In initial velocity ranges where no final velocity is shown, a bion is assumed to form.
  The resonance window where various bounces can be observed has been expanded inside each graphics to better show the fractal pattern. For the sake of comparison the dashed grey line indicates the elastic scenario $v_0=v_f$.}
    \label{Fig:LowAmplitudeViVf}
\end{figure}

\subsubsection{\textit{Velocity diagrams for the scattering between weakly wobbling kinks.}}
In Figure~\ref{Fig:LowAmplitudeViVf}, the velocity diagram for a low initial wobbling amplitude $\widehat{a}_0 = 0.04$ for the orthogonal mode is displayed for different values of $\sigma$. 
Specifically, the values chosen  are $\sigma=1.2$, $\sqrt{7}/2$, $\sqrt{2}$,  $1.5$ and $2.5$, which allows us to describe the entire range of behaviors of the velocity diagrams in this scenario. When the values of $\sigma$ are close to $1$, the resulting fractal structure appears relatively simple, with only a few windows where the wobblers collide more than once. 
The resonant energy transfer mechanism redistributes energy between three modes: two vibrational modes (longitudinal and orthogonal) and the zero mode. The chances of gaining energy by the zero mode are less than in the case where there is only one shape mode. 
Consequently, the number of velocity windows in the resonance interval where the kinetic energy of the wobblers is large enough  to allow them to escape must be smaller compared to the $\phi^4$ model, where only one of these massive modes is present. 
It is important to note that, for the same reason, the one-bounce tail (represented in blue in the upper plots of Figure~\ref{Fig:LowAmplitudeViVf}) exhibits small oscillations, indicating that part of the kinetic energy has been trapped by one of the shape modes.

The previous effect is diminished in the singular case $\sigma_1=\sqrt{7}/2$. As mentioned above, the resonance occurring under the present circumstances  implies that the energy accumulated by the orthogonal shape mode is partially discharged into the longitudinal mode before the collision occurs. 
Scattering can therefore be described in this case as the collision between two longitudinally wobbling kinks. 
Thus, the velocity diagrams found are similar to those observed in the collision between wobblers in the $\phi^4$ model. Note, for example, that the oscillations of the one-bounce tail are more pronounced than in the case $\sigma=1.2$.

As $\sigma$ increases, the fractal structure undergoes substantial changes, becoming increasingly complex. For sufficiently large values of this coupling constant $\sigma$, the diagrams stabilize and resemble the velocity diagram observed for kink/antikink scattering in the $\phi^4$ field theory. 
Furthermore, the critical velocity at which the one-bounce tail emerges is approximately the same, $v_{cr}\approx0.26$ \cite{Campbell1983, AlonsoIzquierdo2021b}. This suggests that when the initial orthogonal amplitude $\widehat{a}_0$ is very small, the shape mode retains its energy without significant transfer to other modes. As a result, the resulting scattering process resembles that of unexcited kinks. 
This phenomenon can be attributed to the fact that the value of $ \widehat{\omega}$ increases as the model parameter $\sigma$ increases. Consequently, exciting the orthogonal mode during collisions becomes much more challenging. Certainly, this behavior changes when the amplitude $\widehat{a}_0$ reaches a certain threshold. 
This suggests that the orthogonal shape mode must be sufficiently excited to initiate the resonant energy transfer mechanism. More discussion on this point will be provided in Section \ref{Sec:section4.2} when the behavior of the amplitude of the internal modes after the collision will be discussed.

Another remarkable phenomenon is that the critical velocity $v_{cr}$, at which the one-bounce tail arises, decreases as $\sigma$ increases. As the energy of the orthogonal mode increases with $\sigma$ this also implies that more energy can be transferred to the zero mode allowing the kink-antinkink pair to escape more easily. As it will be shown in Section \ref{Sec:section4.1.2} and Section \ref{Sec:section4.1.3}, this phenomenon can also be observed for higher values of $\widehat{a}_0$.

\begin{figure}[h!]
\centering
\begin{subfigure}{1\textwidth}
    \includegraphics[width=1.0\linewidth]{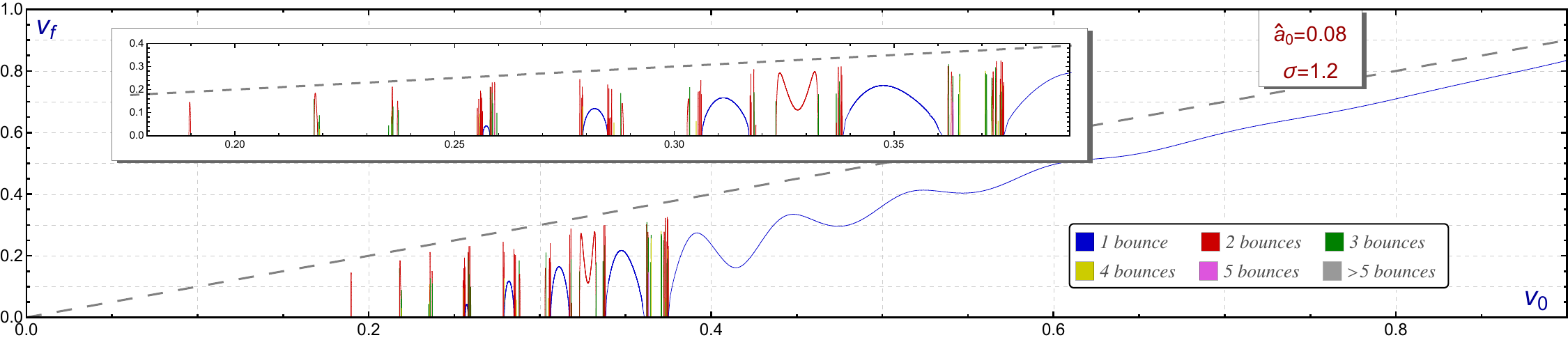}
\end{subfigure}
\hfill
\begin{subfigure}{1\textwidth}    \includegraphics[width=1.0\linewidth]{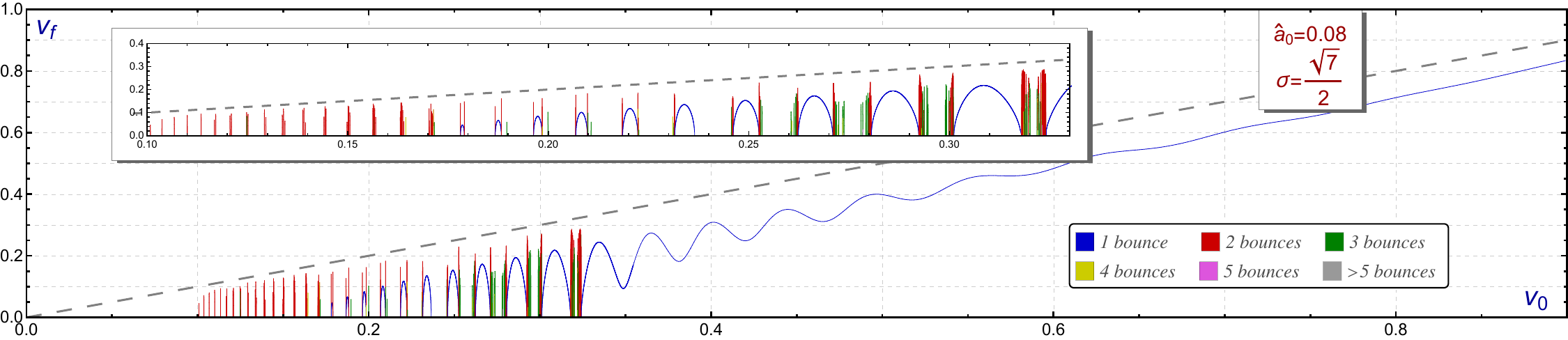}
\end{subfigure}
\hfill
\begin{subfigure}{1\textwidth}
    \includegraphics[width=1.0\linewidth]{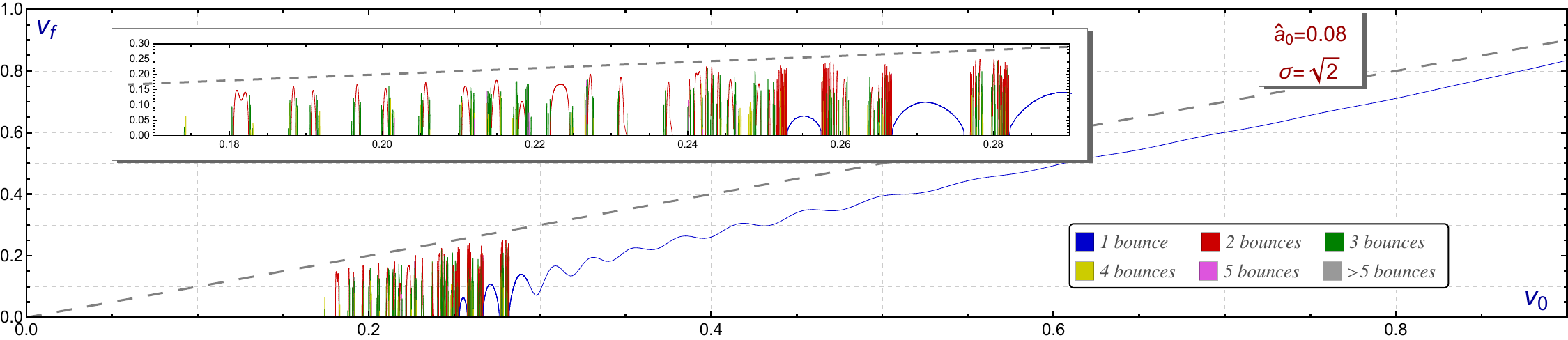}
\end{subfigure}
\hfill
\begin{subfigure}{1\textwidth}
    \includegraphics[width=1.0\linewidth]{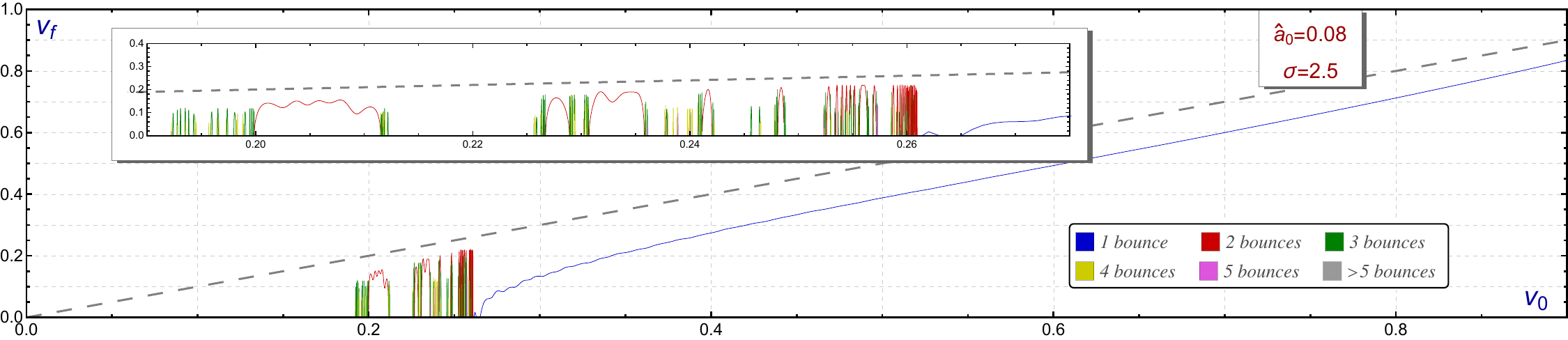}
\end{subfigure}
        
   \caption{Final velocity $v_f$ of the scattered kinks as a function of the initial velocity. The color code shown in the graphs indicates the number of bounces suffered by the kink-antikink pair before moving apart. 
  In initial velocity ranges where no final velocity is shown, a bion is assumed to form.
  The resonance window where various bounces  can be observed has been expanded. For the sake of comparison the dashed grey line indicates the elastic scenario $v_0=v_f$. }
    \label{Fig:VFinalVIncDifferentSgmaA008}
\end{figure}

\subsubsection{\textit{Velocity diagrams for the scattering between moderate wobbling kinks.}} \label{Sec:section4.1.2}
When the initial amplitude of the orthogonal shape mode is increased, the behavior of the velocity diagrams and the fractal pattern that appears in them exhibit novel characteristics with respect to the previous situation, which in turn depends on the value of $\sigma$. 
It is worth noting that in the case of the $\phi^4$ model, the study of wobbling kink collisions led to the identification of new phenomena such as the splitting of the one-bounce tail into isolated one-bounce windows, along with dividing $n$-bounce windows into those with a higher number of bounces. 
These phenomena are also present in the MSTB model with distinctive characteristics. Recall that at the critical value $\sigma_1=\sqrt{7}/2$, the collision of wobblers occurs with a maximum longitudinal wobbling amplitude compared to other $\sigma$ values. 
Consequently, the influence of the orthogonal shape mode decreases, allowing the characteristic features observed in the scattering of wobblers in the $\phi^4$ model to emerge more prominently and distinctively. For other values of $\sigma$, the orthogonal shape mode assumes a more important role in the dynamics. 
For example, when $\sigma<\sigma_1$, the orthogonal frequency is low, which makes the orthogonal shape mode easily excited. 
This implies that the appearance of $n$-bounce windows within the fractal structure  decreases. On the other hand, when $\sigma>\sigma_1$, the number of isolated one-bounce windows decreases, although the complexity of the fractal pattern involving $n$-windows grows and the resonance interval reduces with increasing values of $\sigma$. 
All of these characteristics can be observed in  Figure~\ref{Fig:VFinalVIncDifferentSgmaA008}, where the velocity diagrams are shown for an initial amplitude $\widehat{a}_0=0.08$ for the values of the model parameter $\sigma=1.2$, $\sigma=\sqrt{7}/2$, $\sigma=\sqrt{2}$ and $\sigma=2.5$. 
One of the most notable behaviors occurs in the resonance interval, where the splitting of the $n$-bounce windows into others of the same type occurs, the gap being filled by windows with a greater number of bounces. Additionally, isolated one-bounce windows appear that were not present before. 
Another surprising behavior is that the one-bounce tail begins to exhibit oscillations that, when the initial vibration amplitude is  large enough, lead to the reappearance of isolated one-bounce windows. 
These processes were also present in the scattering of wobbling kinks in the $\phi^4$ model, although, as we will see in the present scenario, these phenomena show novel features, which we describe in more detail in the following points:

\begin{itemize}
\item \textit{Emergence of one-bounce windows in the resonance regime:} 
As a general rule, when $\sigma$ is not close to the critical value $\sigma_1=\sqrt{7}/{2}$, only a few one-bounce windows arising via this mechanism will be observed in the velocity diagrams. In Figure~\ref{Fig:OneBounceWindowSigma1.2}, the velocity plots for one-bounce scattering events are depicted as a function of the initial amplitude of the orthogonal shape mode  for $\sigma=1.2$ within the range $\widehat{a}_0\in[0,0.2]$ using a color-coded scheme. 
In it, red represents the case in which the kinks are not excited, while blue corresponds to the maximum value of the amplitude within the selected range $\widehat{a}_0\in [0,0.20]$. From this figure it is clear that there are only four windows present under the critical velocity  ${v_0 \approx 0.27}$ (associated with unexcited kink-antikink scattering) for relatively high values of $\widehat{a}_0$, and their heights are relatively small. 

\begin{figure}[h!]
    \centering
    \includegraphics[width=\textwidth]{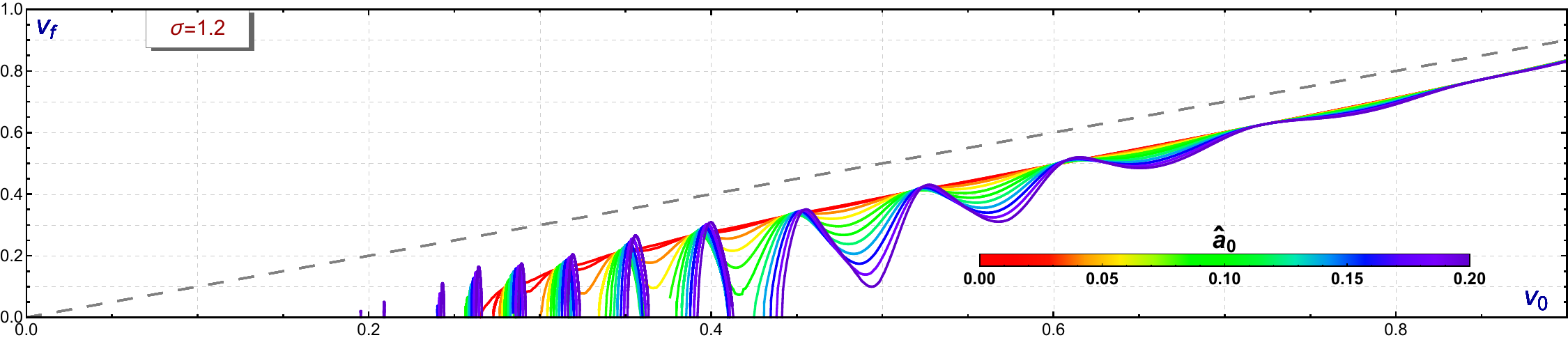}
    \caption{ Velocity diagram for one-bounce scattering events for different values of the initial wobbling amplitude, $\widehat{a}_0\in[0,0.2]$.}
    \label{Fig:OneBounceWindowSigma1.2}
\end{figure}

The presence of these windows can be explained by considering that the larger $\widehat{a}_0$ is, the more energy the kinks possess before the collision. Consequently, more energy can be transferred to the translational mode, allowing the kinks to escape more easily after a collision for relatively low values of $v_0$. This phenomenon can be better understood by referring to Figure~\ref{Fig:ZoomSigma1.2SeveralA}, which represents a sequence of velocity diagrams for closely spaced initial amplitude values $\widehat{a}_0$ for the case of $\sigma=1.2$. 
It can be seen that the fractal structure around $v_0\approx0.255$ is compressed when the amplitude of the orthogonal mode increases. When this amplitude is high enough, a new one-bounce window emerges in the same place where the two-bounce windows existed.

\item\textit{One-bounce reflection tail splitting:} 
As shown in Figures~\ref{Fig:VFinalVIncDifferentSgmaA008} and \ref{Fig:OneBounceWindowSigma1.2}, the one-bounce tail (plotted in blue) begins to oscillate. In fact, for weakly excited wobbling kinks, this phenomenon is already noticeable in the graph associated with $\sigma=1.2$ in Figure~\ref{Fig:LowAmplitudeViVf}. 
With an increase in the amplitude of the orthogonal mode, these oscillations can extend to the $v_f=0$ axis, resulting in isolated one-bounce windows. 
This behavior is more pronounced for values of $\sigma$ close to 1, as seen in Figure~\ref{Fig:VFinalVIncDifferentSgmaA008}. 
This process is clearly illustrated in Figure~\ref{Fig:ZoomSigma1.2SeveralA}, where the one-bounce tail is recurrently split into multiple isolated one-bounce windows. In the sequence represented, one can observe the formation of such a window around $v_0\approx 0.28$ in the second graph, while a new window appears in the third graph for $v_0\approx0.31$. Furthermore, as $\widehat{a}_0$ increases, these new windows begin to shift to the right side of the graph. 
This phenomenon was not observed in the study of scattering between wobblers in the $\phi^4$ model \cite{AlonsoIzquierdo2021b}. 
For example, considering the velocity diagram for $\sigma=1.2$ shown in Figure~\ref{Fig:ZoomSigma1.2SeveralA}, it is evident that the initial point of the second one-bounce window is shifted from $v_0\approx 0.277$ when $\widehat{a}_0=0.04$ to $v_0\approx 0.279$ when $\widehat{a}_0=0.08$. Indeed, this phenomenon is similar to the aforementioned squeezing of the fractal structure that gave as a result new one-bounce windows in the resonant interval of the diagram. The main difference between both phenomena is that now no new one-bounce windows are created; instead, the existing window is compressed to fill the gap between one-bounce windows with $n$-bounce windows. Note in Figure~\ref{Fig:VFinalVIncDifferentSgmaA008} that  splitting  the one-bounce tail results in significantly more isolated one-bounce windows for small values of $\sigma$ than for higher values of $\sigma$.

\begin{figure}[h!]
    \centering
    \includegraphics[trim={3cm 0 3cm 0},width=1.0\linewidth]{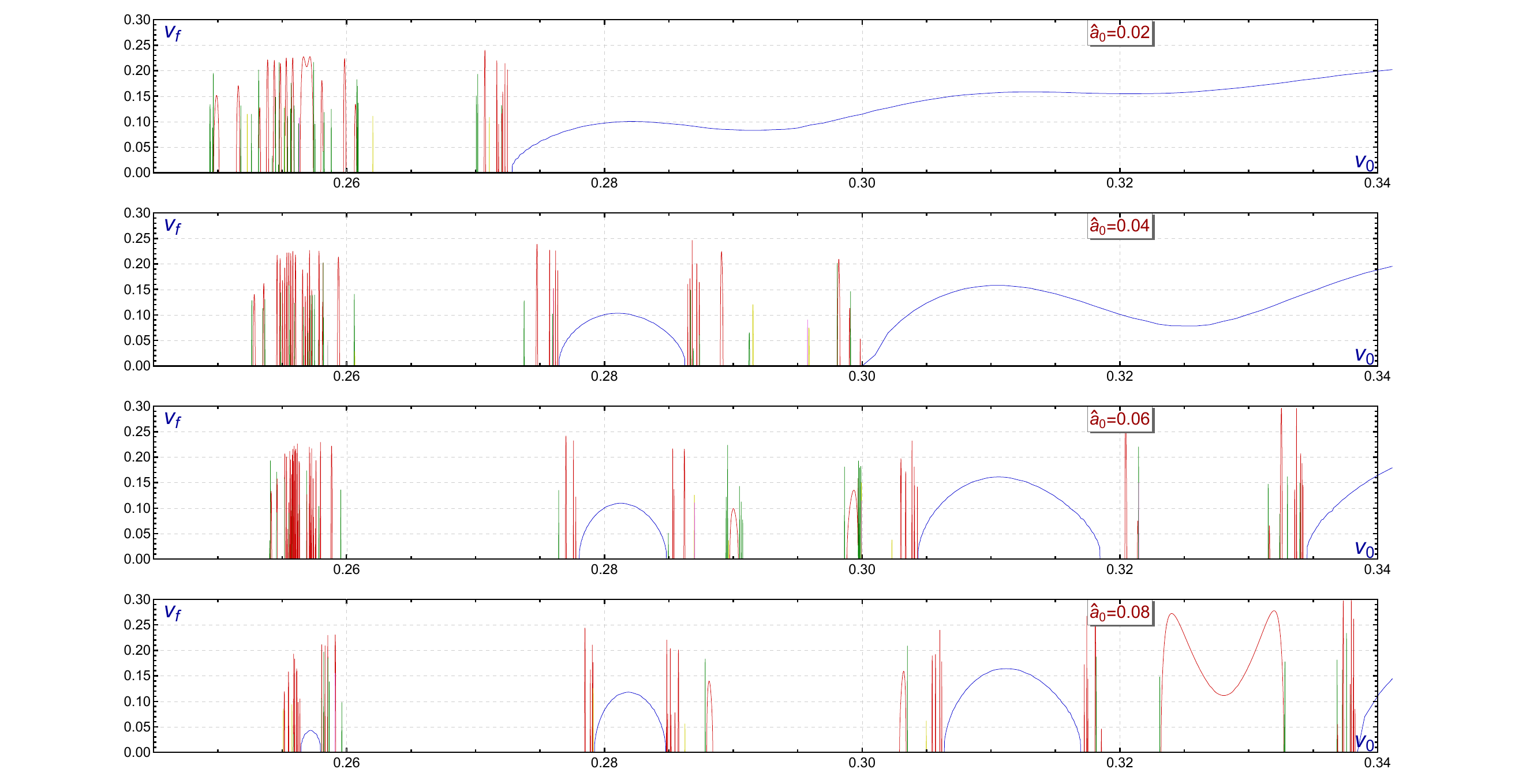}
    \caption{Velocity diagrams for the case $\sigma=1.2$ in the initial velocity range $v_0\in[0.25,0.34]$ for several values of excitation amplitude $\widehat{a}_0$. This graphics illustrates the window splitting mechanism. The color code used to indicate the total number of bounces is the same as in Figure~\ref{Fig:LowAmplitudeViVf}. }
    \label{Fig:ZoomSigma1.2SeveralA}
\end{figure}

\item\textit{$n$-bounce window splitting:} 
Another mechanism that contributes to the increasing complexity of the fractal structure of the velocity diagrams as the initial amplitude increases is the splitting of $n$-bounce windows, where one of these windows is divided into others of the same type. 
This phenomenon already appeared in the $\phi^4$ model.  In this case, increasing the amplitude of the single shape mode results in the generation of more two-bounce windows, which appear progressively from left to right in the corresponding velocity diagram. Figure~\ref{Fig:EvolutionTwoBounce} illustrates the evolution of a single two-bounce window  as $\widehat{a}_0$ goes from $0.02$ to $0.12$ for the kink-antikink scattering in the MSTB model for $\sigma=2.5$. 
It can be seen that as the orthogonal wobbling amplitude increases, a new two-bounce window emerges on the left side of the original window between $\widehat{a}_0=0.06$ and $\widehat{a}_0=0.08$. The gap between these two regions is filled with scattering events exhibiting three and four bounces. 
Subsequently, for $0.10<\widehat{a}_0<0.12$, another window is created, this time on the right side of the plot. This phenomenon can be explained by considering that, for large values of $\sigma$, the frequency of the orthogonal mode is greater than that corresponding to the longitudinal mode. 
Consequently, the vibrational energy of the wobblers increases, providing more energy to transfer to translational mode. This allows the wobbling kink-antikink pair to generate new windows on either side of the original ones as $\widehat{a}_0$ increases.

\begin{figure}[htb]
    \centering
    \includegraphics[width=0.9\linewidth]{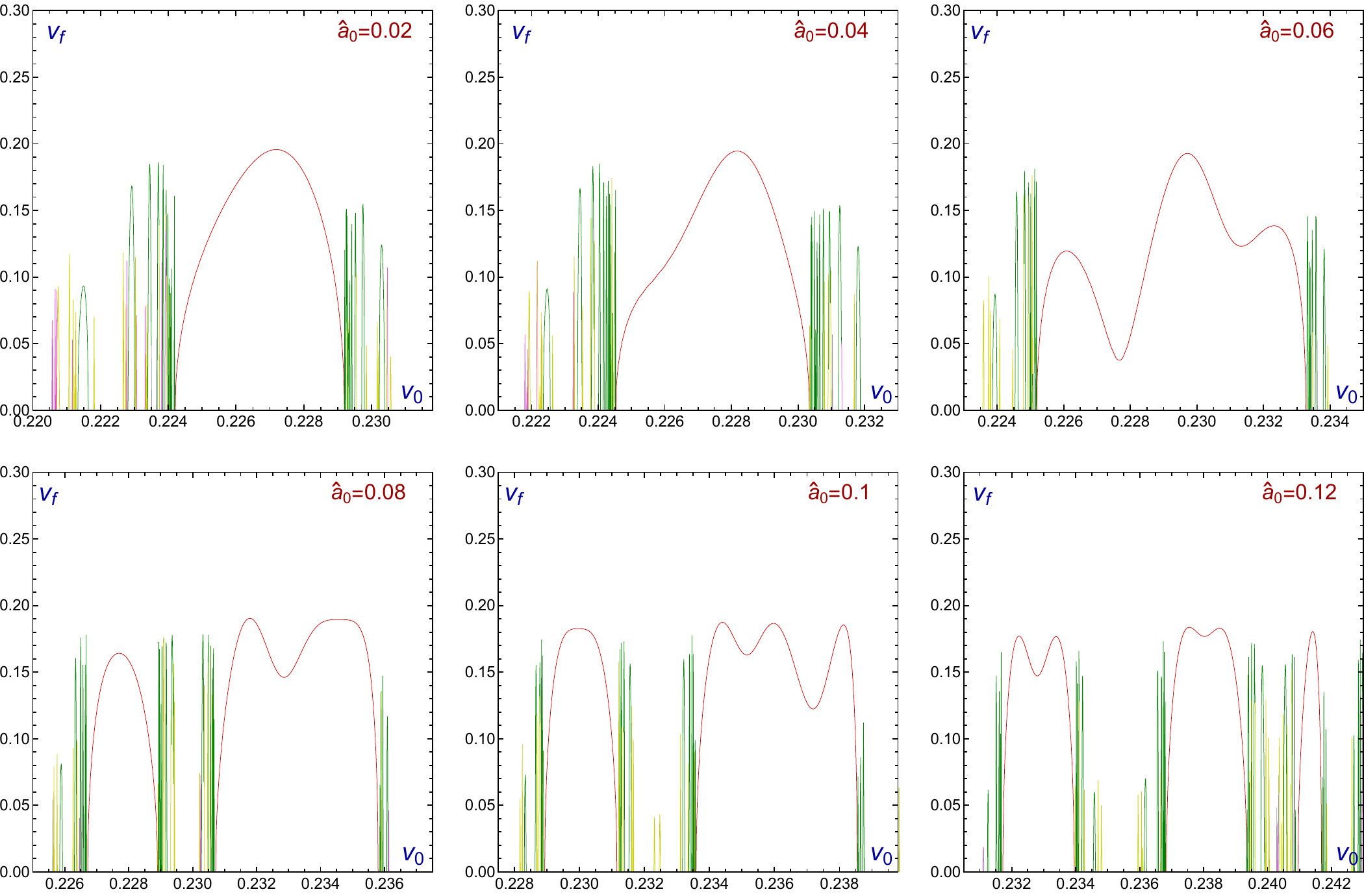}
    \caption{Evolution of a two-bounce window found in the velocity diagram for $\sigma=2.5$ as the value of $\widehat{a}_0$ increases from $\widehat{a}_0=0.02$ to $\widehat{a}_0=0.12$. The color code used to indicate the total number of bounces is the same as in Figure~\ref{Fig:LowAmplitudeViVf}.}
    \label{Fig:EvolutionTwoBounce}
\end{figure}

\end{itemize}

\subsubsection{\textit{Velocity diagrams for the scattering between strongly wobbling kinks.}}  \label{Sec:section4.1.3}

\begin{figure}[h!]
\centering
\begin{subfigure}{1\textwidth}
    \includegraphics[width=1.0\linewidth]{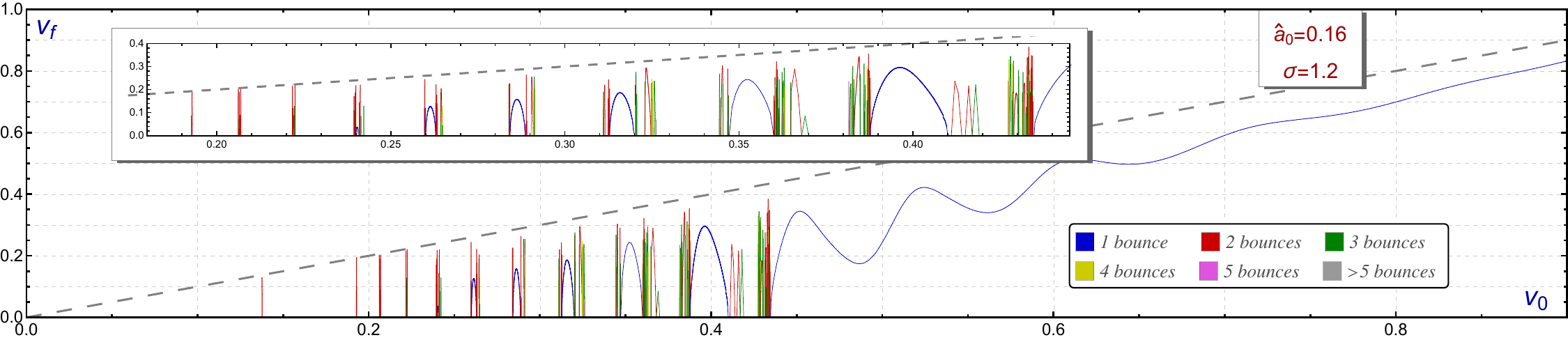}
\end{subfigure}
\hfill
\begin{subfigure}{1\textwidth}
    \includegraphics[width=1.0\linewidth]{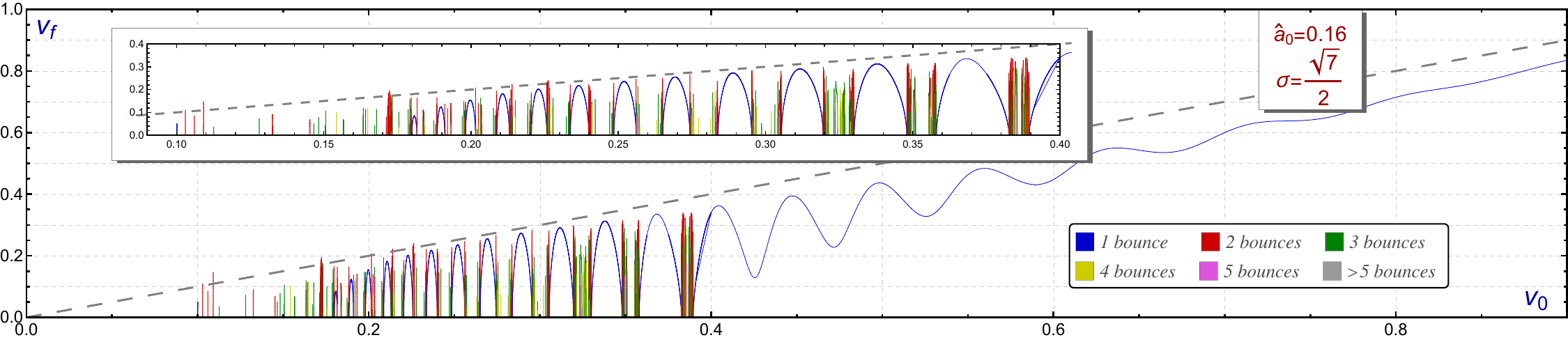}
\end{subfigure}
\hfill
\begin{subfigure}{1\textwidth}
    \includegraphics[width=1.0\linewidth]{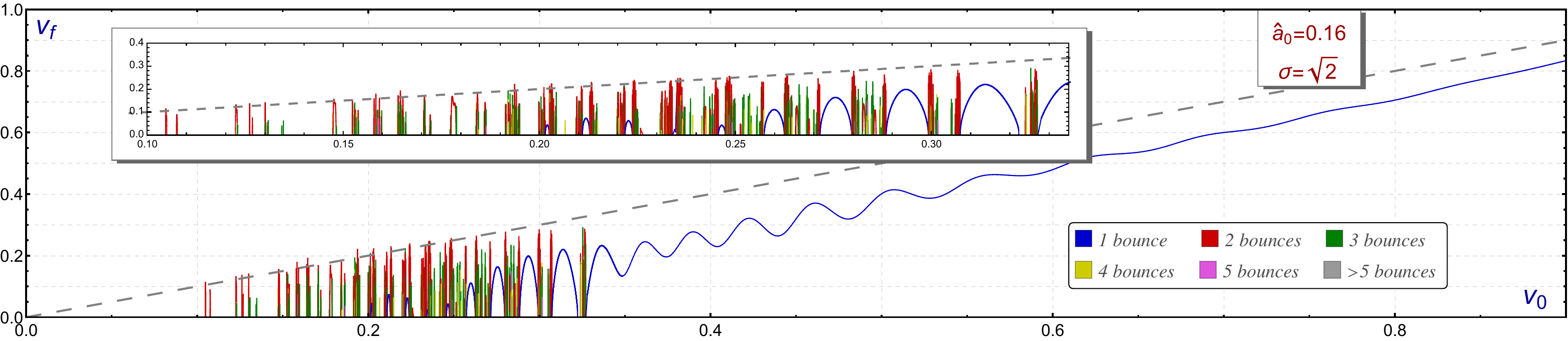}
\end{subfigure}
\hfill
\begin{subfigure}{1\textwidth}
    \includegraphics[width=1.0\linewidth]{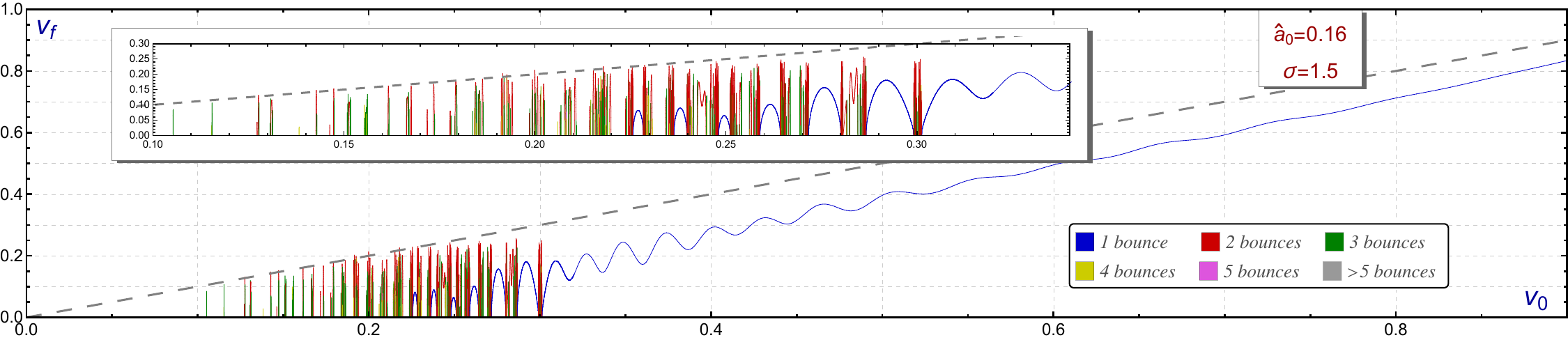}
\end{subfigure}
\hfill
\begin{subfigure}{1\textwidth}
    \includegraphics[width=1.0\linewidth]{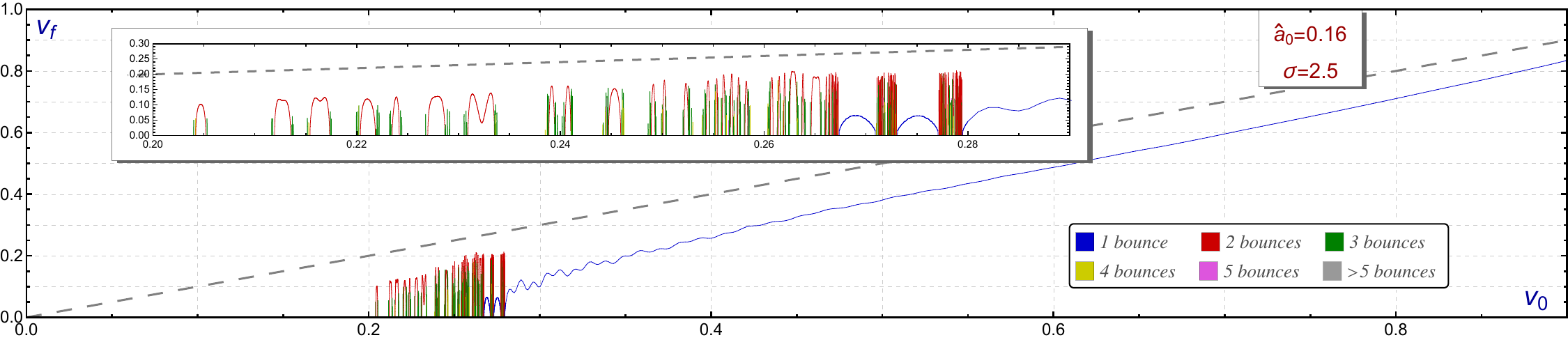}
\end{subfigure}
  \caption{Final velocity $v_f$ of the scattered kinks as a function of the initial velocity. The color code shown in the graphs indicates the number of bounces suffered by the kink-antikink pair before moving apart. In initial velocities ranges where no final velocity is shown, a bion is assumed to form. The resonance window where various bounces can be observed has been expanded. For the sake of comparison tThe dashed grey line indicates the elastic scenario $v_0=v_f$.     }
    \label{Fig:LowAmplitudeViVf16}
\end{figure}

Let us analyze the case where the scattering of orthogonally wobbling kinks occurs with a large initial amplitude $\widehat{a}_0$. In Figure~\ref{Fig:LowAmplitudeViVf16}, the velocity diagrams for the case in which $\widehat{a}_0=0.16$ are depicted. Now, the  behaviors described above are much more pronounced. For example, for small values of $\sigma$, the prevalence of isolated one-bounce windows is observed, which for the case of $\sigma=1.2$ occupies the range between $v_0\in [0.19,0.405]$, and the one-bounce tail presents significant oscillations.

In the singular case $\sigma_1={\sqrt{7}}/{2}$, we observe a highly organized structure in the fractal pattern within the velocity diagram. Isolated one-bounce windows emerge in a repeating pattern, decreasing in both height and width as the initial velocity $v_0$ decreases. As previously mentioned, this is related to the fact that the resonant energy mechanism is practically governed in this case by only two modes (the zero mode and the longitudinal one), resembling the behavior observed in the $\phi^4$ model. It is worth noting that the peak of these windows corresponds closely to the velocity associated with the elastic scenario. 

For the cases where $\sigma=\sqrt{2}$ and $1.5$, the fractal structure is highly complex, also showing isolated one-bounce windows along with others with a greater number of bounces, all densely packed. Note that the regularity seen in the case of $\sigma=\sigma_1$ is completely lost, replaced by a sequence of isolated one-bounce windows that alternate between higher and lower heights. This suggests a higher level of chaotic behavior compared to that observed in the $\phi^4$ model, attributable to the presence of an additional shape mode. 
Similarly, it can be seen in  Figure~\ref{Fig:LowAmplitudeViVf16} how the oscillations of the one-bounce tail decrease as the value of $\sigma$ increases, as seen in the cases of $\sigma=1.5$ and $\sigma=2.5$. Finally, in the case where $\sigma=2.5$, the presence of isolated one-bounce windows is practically suppressed, and the fractal structure is compressed into a much smaller region than that presented in the previously mentioned cases. 
Additionally, it is worth mentioning that the behavior observed in the scattering of wobbling kinks in the $\phi^4$ model, where the kinks transferred a significant amount of energy from the vibrational modes to the zero modes, allowing final velocities of the kinks greater than the initial ones, does not usually appear in the present case. The reason is clearly due to the presence of two shape modes in this case, both accumulating energy, and it is difficult for both to simultaneously transfer their energy to the zero mode to recreate the situation discussed above.

\subsection{Analysis of the shape mode amplitudes} \label{Sec:section4.2}

\begin{figure}[h!]
\centering
\begin{subfigure}{1\textwidth}
    \includegraphics[width=1.0\linewidth]{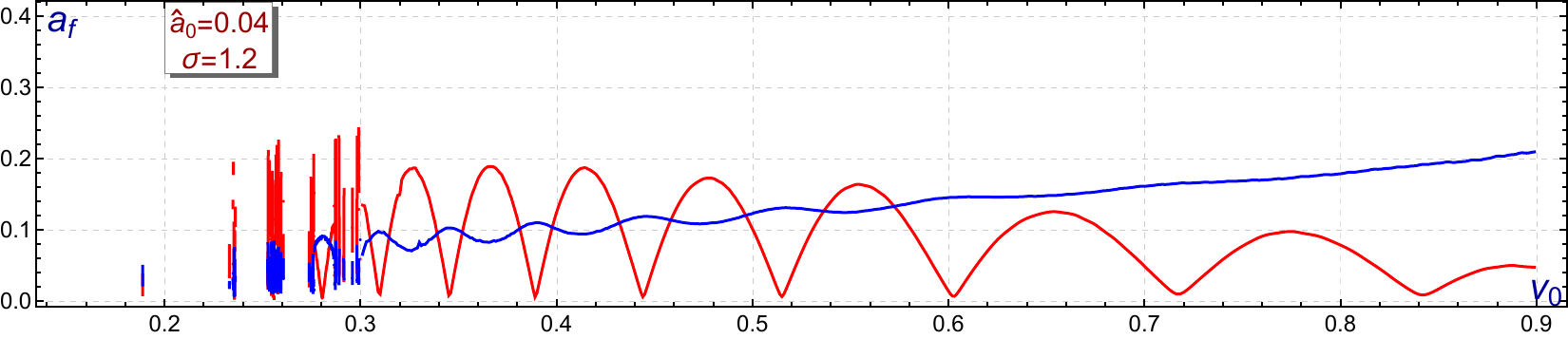}
\end{subfigure}
\hfill
\begin{subfigure}{1\textwidth}
    \includegraphics[width=1.0\linewidth]{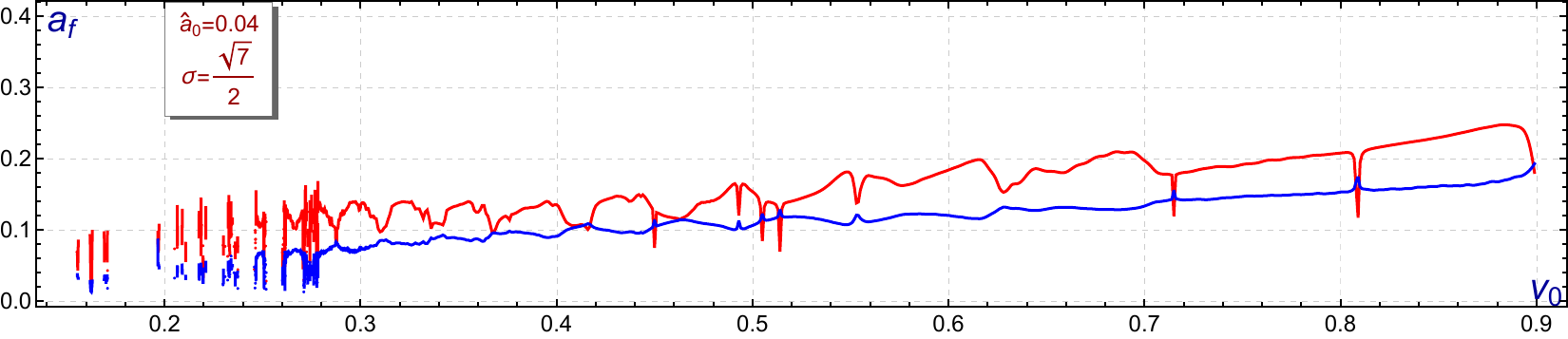}
\end{subfigure}
\hfill
\begin{subfigure}{1\textwidth}
    \includegraphics[width=1.0\linewidth]{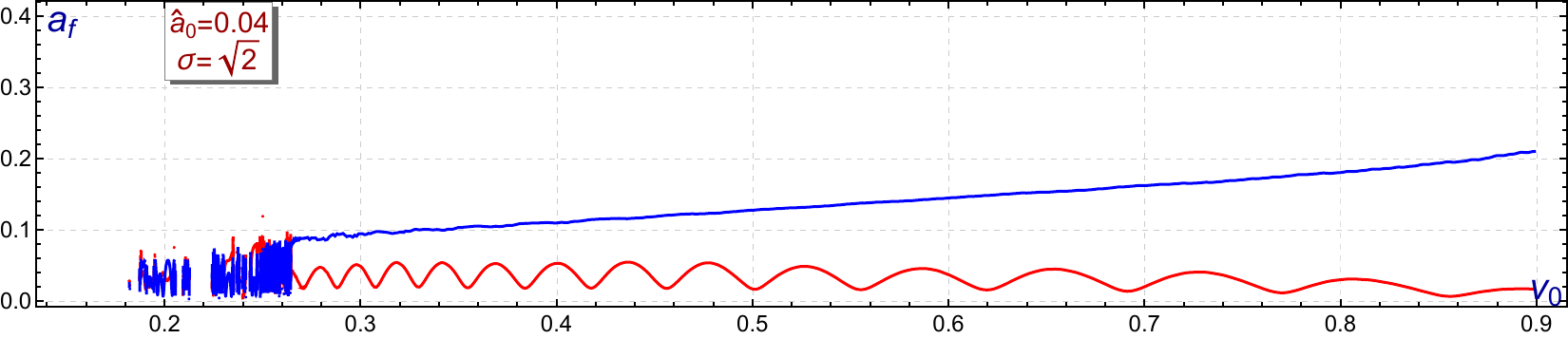}
\end{subfigure}
\hfill
\begin{subfigure}{1\textwidth}
    \includegraphics[width=1.0\linewidth]{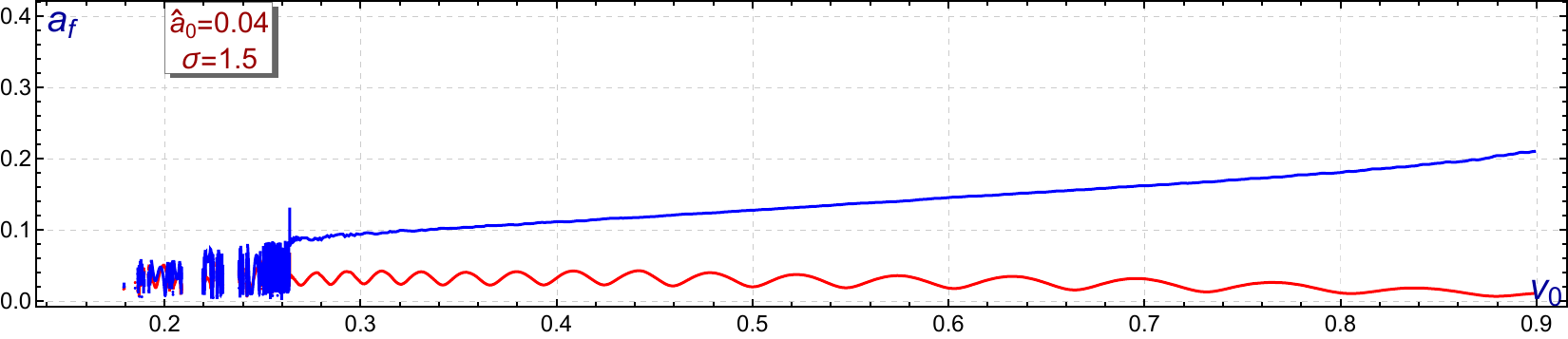}
\end{subfigure}
\hfill
\begin{subfigure}{1\textwidth}
    \includegraphics[width=1.0\linewidth]{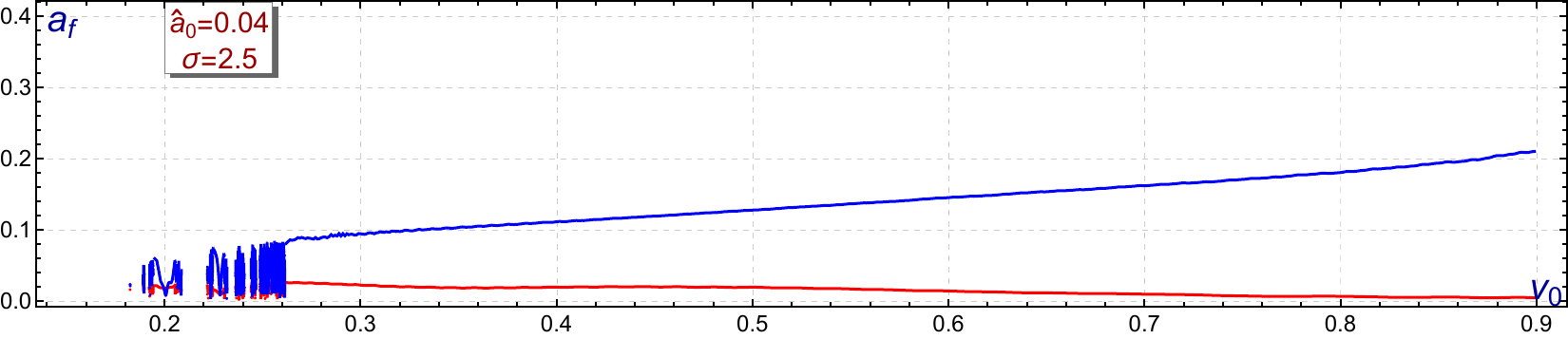}
\end{subfigure}

  \caption{Final wobbling amplitudes of the scattered kinks after the last collision as a function of $v_0$ for various values of $\sigma$ and initial wobbling amplitude $\widehat{a}_0=0.04$. The blue line represents $\overline{a}_f$ (the final longitudinal wobbling amplitude) and the red line represents $\widehat{a}_f$ (the final orthogonal wobbling amplitude).}
    \label{Fig:AmplitudeInternalModesVsViSeveralSigma}
\end{figure}

In this section, we will analyze the behavior of the final vibration amplitudes in both the longitudinal and orthogonal channels. This analysis provides valuable information on how  the resonant energy transfer mechanism works in this model. 
To accomplish this task, the final amplitudes of both shape modes (longitudinal and orthogonal) have been plotted as a function of initial velocity for various values of the amplitude $\widehat{a}_0$ of the orthogonal shape mode and the coupling constant $\sigma$. 
Figure~\ref{Fig:AmplitudeInternalModesVsViSeveralSigma} illustrates the behavior of these amplitudes with the initial amplitude value set to $\widehat{a}_0=0.04$, while the model parameter $\sigma$ takes various values: $\sigma=1.2, \sqrt{7}/2, \sqrt{2}, 1.5$, and $2.5$. 
In the graphs shown in Figure~\ref{Fig:AmplitudeInternalModesVsViSeveralSigma}, the amplitude of the longitudinal shape mode is represented in blue, while that of the orthogonal shape mode is represented in red. The behavior shown in this figure is similar for other values of the amplitude $\widehat{a}_0$. 

It is evident that the dispersion process depends largely on the value of the constant $\sigma$. Analogous to the results obtained previously, we can identify two significant regimes. If the value of $\sigma$ is close to 1 and below the resonance value associated with $\sigma=\sigma_1$, the effect of the resonant energy transfer mechanism becomes particularly pronounced.
There are values of the collision velocity $v_0$ for which the final amplitude of the orthogonal shape mode is almost suppressed, making its value almost disappear. These values are aligned with the peaks of the longitudinal shape mode amplitude. On the other hand, the minima of the longitudinal shape mode amplitude correspond to the maxima of the orthogonal mode amplitude. 
This clearly indicates the energy transfer between these two modes. It is worth noting that the oscillations of the former mode are smaller than those of the latter. This difference arises because the eigenvalue of the orthogonal shape mode is smaller than that of the longitudinal mode, which means that the orthogonal shape mode is less energetic than the longitudinal mode.

\begin{figure}[htb]
      \centering
	   \begin{subfigure}{1\linewidth}
		\includegraphics[width=1.0\linewidth]{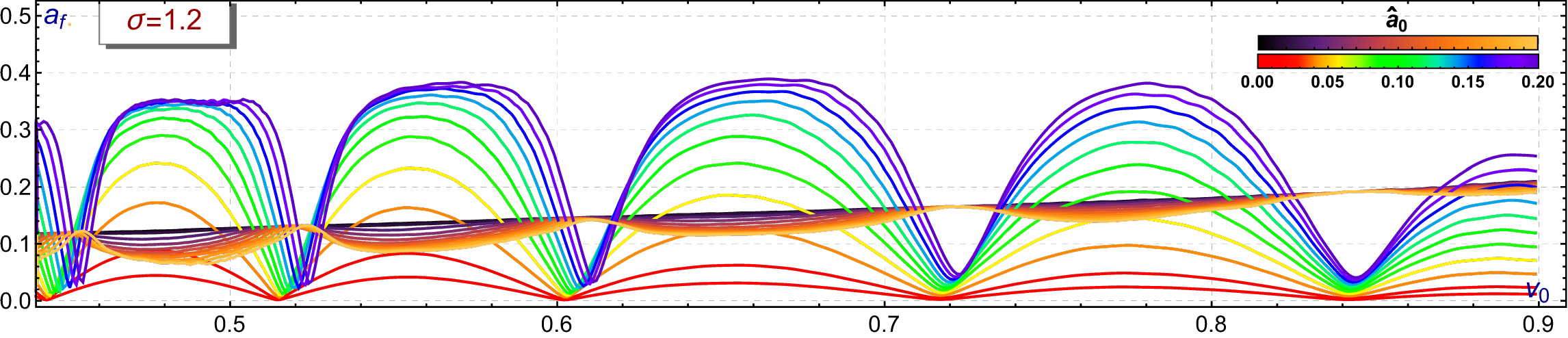}
		\label{Fig:FinalAmplitudesVsV0Sigma12}
	   \end{subfigure}
	   \begin{subfigure}{1\linewidth}
      \vspace{-0.4cm}
		\includegraphics[width=1.0\linewidth]{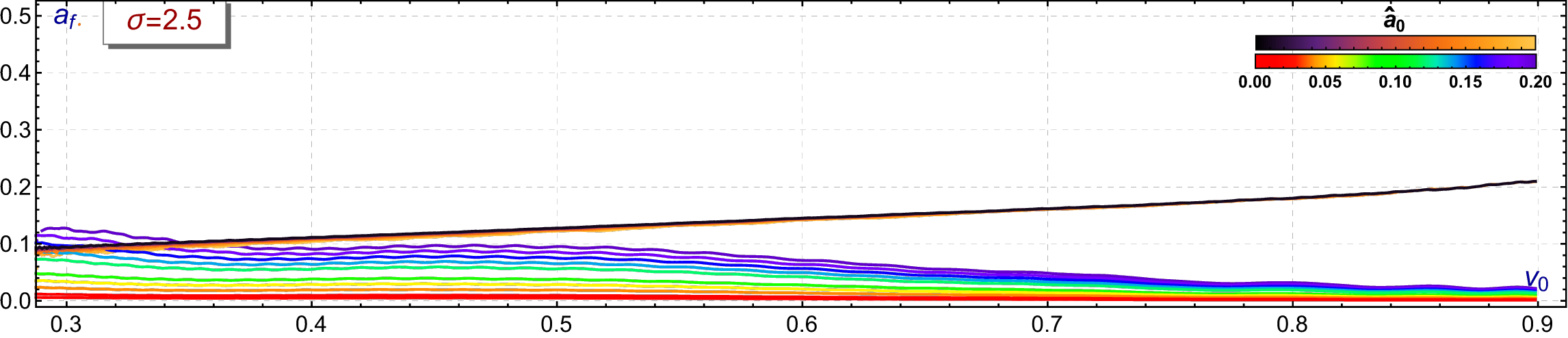}
		\label{Fig:FinalAmplitudesVsV0Sigma25}
	    \end{subfigure}
	\caption{Final amplitudes of both shape modes after the scattering as a function of the initial velocity $v_0$ for various values of $\widehat{a}_0\in [0,0.20]$ and for $\sigma=1.2$ and $2.5$. The first legend located in the upper right corner of the figure corresponds to the amplitude of the longitudinal shape mode while the second corresponds to the amplitude associated with the  orthogonal mode.}
	\label{Fig:FinalAmplitudesVsV0Sigma12-25}
\end{figure}

In the other regime that we mentioned, for values of $\sigma$ greater than $\sigma_1=\sqrt{7}/2$, we observe a similar behavior in the amplitude of the orthogonal shape mode, which oscillates as a function of the initial velocity. However, these oscillations have a smaller amplitude and decrease as the coupling constant $\sigma$ increases. 
In fact, for $\sigma=2.5$, these fluctuations are almost imperceptible. Additionally, it should be noted that, as a general trend, the frequency of these oscillations increases as the value of $\sigma$ increases, which is related to the fact that the frequency of the orthogonal shape mode also increases (see \cite{AlonsoIzquierdo2023}). 
On the other hand, the amplitude of the longitudinal shape mode is not influenced by the orthogonal shape mode. In this case, the excitation of this mode can be attributed almost entirely to the collision of the kinks, or more specifically, to the interaction with the zero modes of these solutions. This behavior is best observed in Figure~\ref{Fig:FinalAmplitudesVsV0Sigma12-25}, where the evolution of the amplitudes corresponding to both shape modes for $\sigma=1.2$ and $\sigma=2.5$ as a function of $v_0$ and $\widehat{a}_0$ can be observed. 
As shown in the top graph of Figure~\ref{Fig:FinalAmplitudesVsV0Sigma12-25}, for low values of $\sigma$ increasing the value of the initial amplitude $\widehat{a}_0$ results in a greater  final amplitude of orthogonal shape mode $\widehat{\eta}$ after the last collision. 
Simultaneously, as the maxima corresponding to $\widehat{a}_f$ increase, the minima of $\overline{a}_f$ decrease. Note that there is an energy loss in the longitudinal shape mode compared to the case where the orthogonal shape mode is not initially excited. 
Another important aspect to highlight is that the nodes of the oscillations exhibited by the shape modes depend on the initial amplitude. As seen in Figure~\ref{Fig:FinalAmplitudesVsV0Sigma12-25}, the nodes shift slightly to the right as the value of $\widehat{a}_0$ increases.
On the other hand, as shown in the lower graph of Figure~\ref{Fig:FinalAmplitudesVsV0Sigma12-25}, the behavior of the longitudinal shape mode amplitude does not depend on the initial wobbling amplitude $\widehat{a}_0$ when $\sigma$ is high enough. In fact, the wobbling amplitude of the orthogonal mode stops oscillating and its value decreases as the kinks collide with greater velocity.

\begin{figure}[htb]
\centering
\begin{subfigure}{1\textwidth}
    \includegraphics[width=1.0\linewidth]{Section4.2/AmplitudeInternalModesVsViS13A004.pdf}
\end{subfigure}
\hfill
\begin{subfigure}{1\textwidth}
    \includegraphics[width=1.0\linewidth]{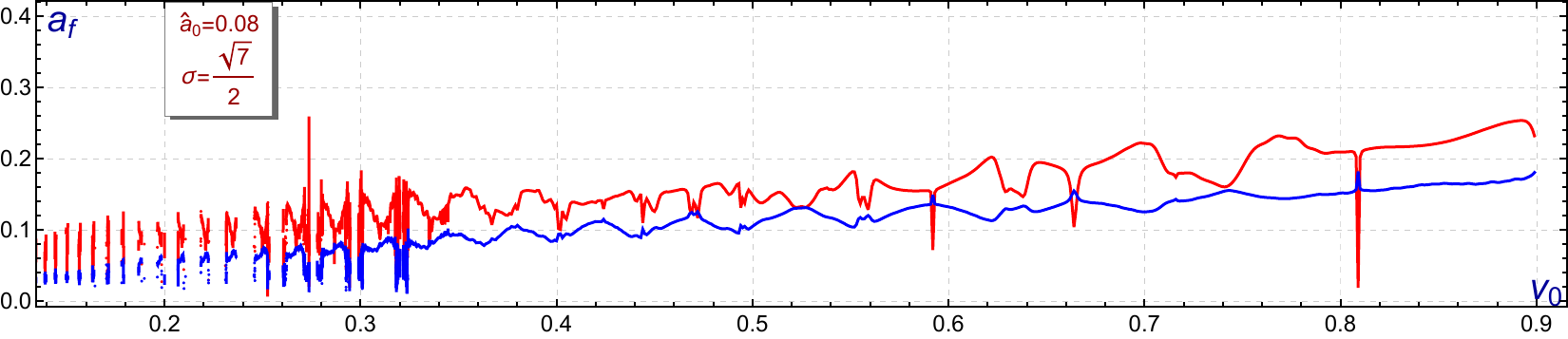}
\end{subfigure}
\hfill
\begin{subfigure}{1\textwidth}
    \includegraphics[width=1.0\linewidth]{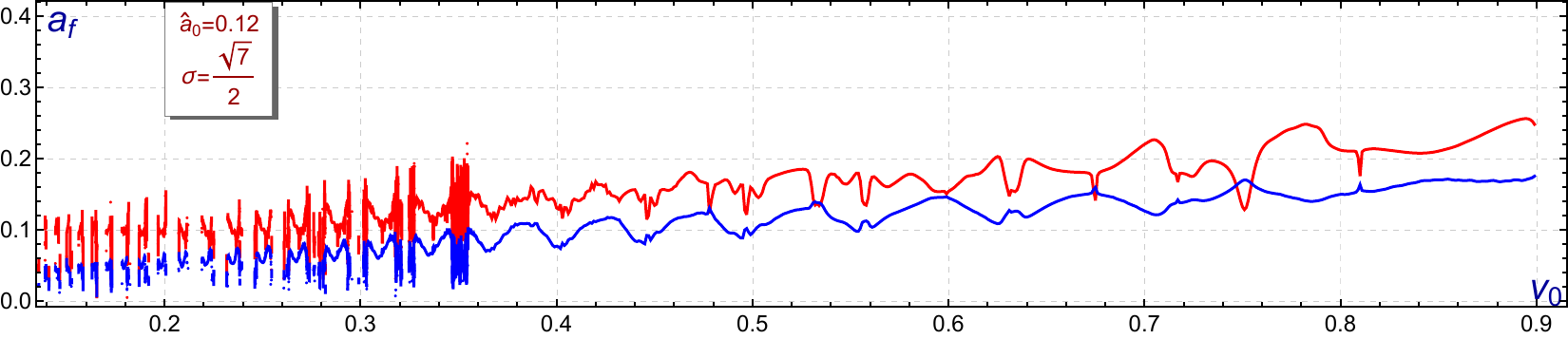}
\end{subfigure}
\caption{Final wobbling amplitudes for the scattered kinks after the last collision as a function of $v_0$ for $\sigma=\sqrt{7}/2$ and wobbling amplitude $\widehat{a}_0=0.04$. The blue line represents $\overline{a}_f$ (the final longitudinal wobbling amplitude) and the red line represents $\widehat{a}_f$ (the final orthogonal wobbling amplitude).}
    \label{Fig:AmplitudeInternalModesVsViSeveralA0S13}
\end{figure}

However, the behavior of the system is completely different for the special value $\sigma_1=\sqrt{7}/2$, where a resonance occurs between the frequencies $\overline{\omega}$ and $2\widehat{\omega}$. In this case, the graphical representation of the amplitudes is more irregular than in the other cases and, surprisingly, the final amplitude of both shape modes does not depend on the initial wobbling amplitude $\widehat{a}_0$, being $\widehat{a}_f$ almost always slightly larger than the amplitude corresponding to the longitudinal eigenmode (see Figure~\ref{Fig:AmplitudeInternalModesVsViSeveralA0S13}). 
The mechanism that seems to explain the scattering of the initially excited orthogonally wobbling kinks in this case is the following: when these kinks approach each other, a significant energy transfer from the orthogonal shape mode to the longitudinal mode occurs due to the aforementioned resonance \cite{AlonsoIzquierdo2023}. 
This results in the orthogonal shape mode being weakly excited immediately before the collision of the wobblers, while the longitudinal mode has gained considerable energy. This leads, for example, to velocity diagrams similar to those found in \cite{AlonsoIzquierdo2021b} for the $\phi^4$ model, where there is no second component. 
After the collision, the longitudinal mode is assumed to be quite excited, but a channel exists that recharges the orthogonal mode through higher-order nonlinear couplings. Since both modes have been excited, the energy transfer channel from the orthogonal to the longitudinal mode is suppressed.
However, it should be noted that for this critical value of $\sigma$, there is also strong radiation emission in the orthogonal channel with frequencies $\omega=\overline{\omega}+ \widehat{\omega}$ and $3\widehat{\omega}$, resulting in an increase in decay in the shape mode chain. It is important to consider that this process is quite complex, and its description through numerical schemes may depend on the $t_{\rm max}$ used in the simulations, which justifies the irregular behavior of the graphs shown for this value. 
Furthermore, the reason behind $\widehat{a}_f>\overline{a}_f$ can be explained by considering that the longitudinal eigenvalue is larger than the orthogonal one, which makes the orthogonal mode easier to excite. In fact, it can be seen in Figure~\ref{Fig:AmplitudeInternalModesVsViSeveralA0S13} that in the one bounce tail regime $\frac{\overline{a}_f}{\widehat{a}_f}\approx\frac{\overline{\omega}}{\widehat{\omega}}$.

\section{Concluding remarks}\label{Sec:section5}

In this paper, we have investigated the interaction between the translational mode and the shape modes in a two-component scalar field theory in the context of kink-antikink scattering processes when only the mode corresponding to the second field component is initially activated. 
The energy transfer mechanism between modes has already been discussed, both analytically and numerically in \cite{AlonsoIzquierdo2023}, but only for a static kink whose orthogonal mode had initially been  triggered. The results presented in this work are therefore a natural continuation of the aforementioned paper, shedding light on how the resonant energy mechanism  affects the fractal structure found in the velocity diagrams. 
The presence of two different shape modes makes the energy transfer mechanism more complex than when wobbler collisions were studied in the $\phi^4$ model. Since there are now two different discrete eigenmodes instead of just one, there is less  kinetic energy available for the translational mode, which explains why the escape velocity of the kink-antikink pair after the last collision $v_f$ is never greater than the initial kink velocity collision $v_0$. This is in contrasts to what was found in the $\phi^4$ model, since in the latter, when the amplitude of the only shape mode was triggered with enough amplitude, $v_f$ could be larger than $v_0$.

Another notable phenomenon found in this study is the changing of the velocity diagrams as $\widehat{a}_0$ and $\sigma$ change their value. For low values of $\sigma$, when $\widehat{a}_0$ increases, one-bounce windows were created in the resonant part of the velocity diagram. 
However, for high values of $\sigma$ and a small value for $\widehat{a}_0$ the velocity diagram resembles the behavior found for an unexcited kink in the $\phi^4$ field theory. This phenomenon does not manifest itself for higher values of the initial orthogonal amplitude, since in the latter case the resonant part of the diagram becomes chaotic and the amplitude of the  oscillations of the one-bounce regime decreases.    

The system response changes dramatically  when the value of $\sigma$ is set to $\sqrt{7}/2$. Under these conditions, the orthogonal mode discharges most of its energy into the longitudinal mode and  radiation modes. This implies that the longitudinal mode is activated  with a certain amplitude before the scattering process. Such circumstance clarifies the similarity between the velocity diagrams for the resonance value $\sigma_1=\sqrt{7}/2$ and the velocity diagrams coming from the wobbler scattering in the $\phi^4$ model \cite{AlonsoIzquierdo2021b,AlonsoIzquierdo2022}.

In addition to what has already been explained, it is worth highlighting the correspondence between the maxima and minima of the final wobbling amplitudes explained in detail in Section~\ref{Sec:section4.2}, which has shed light on the energy transfer mechanism between modes which, as has being shown, strongly depends on the value of $\sigma$ since the orthogonal mode can only be triggered again after the scattering process for values of $\sigma\approx 1$.

As a future line of research that serves as a natural continuation of this work, we consider extending the numerical techniques presented here in order to study the scattering of excited topological defects defined in higher dimensions. Among the innumerable possibilities that are contemplated in this regard, one of them is generalize the results found for the excited vortex scattering process in the Abelian-Higgs model recently presented  in  \cite{AlonsoIzquierdo2024b, Krusch2024} studying the same situation in generalized models that support the existence of vortices. Another potential avenue of exploration considered involves the study  of  kink-antikink  collisions in two-component scalar field models whose kink solutions possess more than three internal modes. In this regard, a suitable candidate for this research would be the two-component coupled $\phi^4$ model \cite{Halavanau2012, AlonsoIzquierdo2024}, in which the internal structure and the possible number of eigenmodes depend on the value of the coupling constant that the potential defining the model depends on.

\section*{Acknowledgments}

This research was supported by the European Union-Next Generation UE/MICIU/Plan de Recuperacion, Transformacion y Resiliencia/Junta de Castilla y Leon (PRTRC17.11), and also by RED2022-134301-T and PID2020-113406GB-I00, both financed by MICIU/AEI/10.13039/501100011033.
DMC acknowledges financial support from the European Social Fund, the Operational Program of Junta de Castilla y Leon and the regional Ministry of Education.
This research has made use of the high-performance computing resources of the Castilla y Le\'on Supercomputing Center (SCAYLE), financed by the European Regional Development Fund (ERDF).


\end{document}